\documentclass[11pt, preprint]{aastex}

\usepackage{geometry}                
\usepackage{graphicx}
\usepackage{amssymb}
\usepackage{epstopdf}
\usepackage{longtable}
\newcommand{\Msun}{\hbox{M$\sb{\odot}$}}
\newcommand{\Rsun}{\hbox{R$\sb{\odot}$}}

\begin{document}

\title{Chemical Abundances in the Externally Polluted White Dwarf GD 40: \\Evidence of a Rocky Extrasolar Minor Planet}

\author{B. Klein\altaffilmark{1}, M. Jura\altaffilmark{1}, D. Koester\altaffilmark{2}, B. Zuckerman\altaffilmark{1}, and C. Melis\altaffilmark{1,3}} 

\altaffiltext{1}{Department of Physics and Astronomy, University of California, Los Angeles, CA 90095-1562, USA; kleinb, jura, ben@astro.ucla.edu}
\altaffiltext{2}{Institut fur Theoretische Physik und Astrophysik, University of Kiel, 24098 Kiel, Germany; koester@astrophysik.uni-kiel.de}
\altaffiltext{3}{Current address: Center for Astrophysics and Space Sciences, University of California, 
San Diego, CA 92093-0424, USA; cmelis@ucsd.edu }

\begin{abstract}
We present Keck/HIRES data with model atmosphere analysis of the helium-dominated polluted white dwarf GD 40, in which we measure atmospheric abundances relative to helium of 9 elements:  H, O, Mg, Si, Ca, Ti, Cr, Mn, and Fe.  Apart from hydrogen whose association with the other contaminants is uncertain, this material most likely accreted from GD 40's circumstellar dust disk whose existence is demonstrated by excess infrared emission.  The data are best explained by accretion of rocky planetary material, in which heavy elements are largely contained within oxides, derived from a tidally disrupted minor planet at least the mass of Juno, and probably as massive as Vesta.  The relatively low hydrogen abundance sets an upper limit of 10\% water by mass in the inferred parent body, and the relatively high abundances of refractory elements, Ca and Ti, may indicate high-temperature processing.  While the overall constitution of the parent body is similar to the bulk Earth being over 85\% by mass composed of oxygen, magnesium, silicon and iron, we find n(Si)/n(Mg) = 0.30 $\pm$ 0.11, significantly smaller than the ratio near unity for the bulk Earth, chondrites, the Sun, and nearby stars.  This result suggests that differentiation occurred within the parent body.
\end{abstract}


\maketitle

\section{INTRODUCTION}

Detailed information about the bulk compositions of planetary bodies is generally limited by the lack of available mechanisms to probe planet interiors.  However, the technique of measuring accreted element abundances of tidally disrupted planetary bodies that pollute the atmospheres of cool white dwarfs provides an indirect, yet sensitive method to study bulk compositions.  The otherwise pure white dwarf atmospheres are a clear canvas for displaying the relative bulk elemental abundances of the swallowed planetary parent body/bodies, as demonstrated with Keck/HIRES observations, and subsequent abundance analyses, of the dramatically polluted white dwarf GD 362 (Zuckerman et al. 2007, Koester 2009a, Jura et al. 2009a).

In contrast to main sequence stars whose atmospheres contain mainly primordial elements, cool (6,000 K $<$ $T_{eff}$ $<$ 25,000 K) white dwarfs provide a relatively clean slate since their high surface gravities cause heavy elements to settle downward, leaving above a nearly pure hydrogen and/or helium atmosphere (Schatzman 1945).  The settling times for ``high-Z''  (atomic \#Z $>$ 2) elements to diffuse down from the outer convection zone are always short compared to the white dwarf lifetime (Paquette et al. 1986), and thus the high-Z elements cannot be primordial.  The competing outward force of radiative levitation is negligible below 25,000 K (Chayer et al. 1995), and convective carbon dredge-up, which may play a role in bringing carbon to the surface for $T_{eff}$ $<$ 13,000 K (Koester et al. 1982), cannot explain the presence of heavier elements.  Consequently, high-Z atmospheric contaminants in cool white dwarfs must be accreted from an external source.   Given the purity of white dwarf (WD) atmospheres, even trace amounts of external pollutants can be detected spectroscopically, especially with high resolution, high signal-to-noise ratio (SNR) observations.

While heavy element absorption features have been detected in spectra of single WDs for over half a century (Kuiper 1941; see also Sion et al.$\;$1983; Wesemael et al.$\;$1993 for current WD classifications and references), it is relatively recent that the growing number of discoveries of infrared excess emission (due to light-reprocessing by the dust of a circumstellar disk) at many of the most heavily polluted WDs has elucidated the nature of the accretion reservoirs (Zuckerman \& Becklin 1987; Becklin et al. 2005; Kilic et al. 2005, 2006; von Hippel et al. 2007; Jura et al. 2007a, 2007b; Brinkworth et al. 2009; Farihi et al. 2009; Melis et al. 2009, in preparation).  The spectral energy distribution of the infrared excess is well modeled by a disk geometry and the dust temperatures typically range from inner to outer edges between 1200 K and 300 K (Jura et al. 2007a), which correspond to radii between 10 and 100 R$_{*}$ (0.13 - 1.5 $\Rsun$), within the tidal radius of a typical white dwarf (Jura 2003).   Double peaked Ca {\small II} emission line profiles at some of the hotter ($T_{eff}$ near 20,000 K) WDs reveal the dynamics of gaseous material in Keplerian rotation also within 1.5$\;\Rsun$ (Gaensicke et al. 2006, 2007, 2008).  Given the location, the dust and gas must have been produced (or delivered) after the stars' red giant phase.  Additional insight into the nature of the circumstellar material comes from the 10$\micron$ silicate emission features detected with $\it{Spitzer}$ IRS spectra of dusty white dwarfs (Jura et al. 2007b, 2009b, Reach et al. 2005, 2009), which can be modeled by micron-sized olivine-like glasses that produce a characteristic emission profile resembling that of circumstellar, but not interstellar grains.
Evidence is strong that these disks are derived from the tidal disruption of an asteroid or minor planet (see Jura 2008) which has undergone orbital perturbation (Debes \& Sigurdsson 2002) during post-red giant evolution of the star's planetary system. 

 The material in the atmospheres of cool white dwarfs with detected disks has almost certainly been accreted from their circumstellar reservoirs.  The atmospheric contaminant abundances are linked to the composition of the accreted source material (Koester 2009a).  This is potentially a rich source of highly detailed information regarding the nature of extrasolar planetary matter.  In their work on GD 362 with Keck/HIRES, Zuckerman et al. (2007) detected 15 high-Z elements, a major advance over the low resolution discovery work that detected Ca, Mg, and Fe (Gianninas et al. 2004).  Zuckerman et al. (2007) found element abundances consistent with a refractory-rich, volatile-poor composition which they conclude very probably derived from an asteroid or minor planet.  What is needed are more high-spectral-resolution, high-SNR observations and abundance analyses of additional WD/perturbed-planetary systems.  In this paper we present a detailed analysis of the atmospheric composition of GD 40 (= WD 0300-013), an exemplar of polluted and dusty white dwarfs.  For the first time, ground-based observations of an externally polluted WD have detected all the major constituents of terrestrial planets: O, Mg, Si, and Fe.  It is a useful reminder that these 4 elements compose 94\% of Earth's mass.  With current models and calculations of the relationship between accreted and atmospheric abundances in polluted white dwarfs (Koester 2009a, Jura et al.$\;$2009a), we focus on understanding the bulk chemical composition of GD 40's polluting parent body.  We show that rocky extrasolar planetary bodies exist(ed) in this presumably ancient planetary system and may have experienced differentiation during formation and/or evolution.

\section{OBSERVATIONS \& MEASUREMENTS}

The data were acquired in good weather conditions with the high resolution echelle spectrometer, HIRES (Vogt et al. 1994) on the Keck I telescope at Mauna Kea Observatory.  A 1.148 arcsecond slit width was used for all observations, providing a spectral resolution of R = $\lambda$/$\Delta\lambda$ $\sim$ 40,000.  Table 1 lists observation dates and setups.

GD 40 (V$_{mag}$ = 15.56, spectral type DBAZ) is the prototype of the DBZ class, being the first helium-lined cool white dwarf observed to display heavy element traces  (Wickramasinghe et al. 1975, Shipman \& Greenstein 1983).  HST FOS observations in the UV (Shipman et al 1995), and the subsequent analysis by Friedrich et al. (1999) and Wolff et al. (2002), hereafter F99W02, yielded abundance measurements of H, C, Mg, Si, Ca, and Fe, with uncertainties $\sim$ 0.5 dex.  An optical spectrum of GD 40 (Koester et al. 2005a) was obtained in the ESO Supernova Progenitor (SPY) survey, and has been used in determining the effective atmospheric temperature, T$_{eff}$ = 15,300 K, with log $g$ set to 8.0 (Voss et al. 2007), which we adopt for the models used in this work ($\S$ 3 contains further discussion of T$_{eff}$ and log $g$).  Recently, GD 40 was found to possess an infrared excess (Jura et al. 2007a) and 10$\micron$ silicate emission of circumstellar origin (Jura et al. 2009b).  These discoveries made Keck/HIRES observations of GD 40 a priority.

\subsection{Data Reduction}

Data reduction was carried out with standard IRAF\footnotemark[1]\footnotetext[1]{http://iraf.noao.edu/docs/recommend.html; Churchill (1995) is also instructive} tasks for echelle data.  A PyRAF code meta-structure controls the data processing flow, maintains file organization and adds versatility.  The data have undergone typical echelle reduction procedures:  bias removal, flat-fielding, variance weighted (``optimal") extraction, dispersion calibration with linearization and heliocentric doppler correction.  Much of the echelle blaze function was removed by dividing each spectrum by a low-order polynomial fit to the extracted spectrum of the quartz lamp flat exposure.  It was desirable to increase the signal-to-noise ratio by combining orders of overlapping wavelength coverage, and to reduce the known HIRES artifacts (waves and bends in the continuum of a wavelength scale greater than $\sim$10 \AA) most likely due to variable vignetting (Suzuki et al. 2003).  Since we do not require absolute flux calibrated information, and we are mainly interested in heavy-element absorption lines (equivalent widths $<$ 3 \AA), the echelle orders were continuum normalized to unity with a low-order polynomial (typically 5th order cubic spline) to reduce artifacts and align the echelle orders.  An error associated with this action was estimated by varying the order of the normalization polynomial, which resulted in a $\lesssim$ 1\% variation in equivalent width measurements  of typical heavy element absorption lines.   Adjacent orders of overlapping wavelength coverage were combined in an average, weighted by the blaze function to suppress the contribution from the noisier edges of the CCD chip (e.g. Aoki 2006\footnotemark[2]\footnotetext[2]{http://optik2.mtk.nao.ac.jp/$\sim$waoki/QL/specana200601e.pdf}), and a final order-merged, normalized spectrum was obtained.   

We employed different instrumental setups to achieve our desired wavelength coverage and sensitivity.  For the 2008 red data, an additional re-normalizing processing step had to be applied to calibrate and remove 2nd (diffraction) order blue flux contamination in the region $\lambda$8200-9000{\AA}.  The 2006 red data had a different instrumental setup, providing an un-contaminated comparison to check our renormalization procedure.  The final all-combined GD 40 HIRES spectrum has SNRs in the range of 35 to 50 for the spectral interval 3300 - 4500{\AA}, and between 60 - 90 for 4500 - 7000\AA. $\;$Shorter than 3300\AA, the SNR is $\simeq$ 20;  near 7000\AA, the SNR is $\simeq$ 50; redward of 8000\AA, the SNR is $\leq$ 35.

\subsection{Spectral Lines}
GD 40's optical spectrum is dominated by large (equivalent widths $\sim$ 1-10 \AA) pressure broadened  helium lines, a signature of the spectral class ``DB''.  The high-resolution and high-SNR Keck/HIRES spectrum displays absorption lines of many more species.  In addition to He we identify the presence of 9 elements, H, O, Mg, Si, Ca, Ti, Cr, Mn, and Fe (Figures 1 -- 8), whose detections are securely established through multiple ($\geq$ 4), unambiguous lines for all elements heavier than hydrogen.  Hydrogen is detected via a weak but clearly visible H$\alpha$ feature, also observed by Voss et al. (2007).   Except for the Ca {\small II} H\&K lines and two Ti {\small II} lines (3242.0 \AA$\;$and 3383.8 \AA), GD 40's absorption lines arise from excitation levels above the ground state, and are therefore not of interstellar origin. The very large and broad profiles of the Ca {\small II} H\&K lines indicate their atmospheric origin.  The carbon detection from UV data (Shipman et al. 1995, F99W02) brings the number of known elements aside from He in GD 40's atmosphere to 10.

Each absorption line was visually inspected.  Line identifications made use of the Kurucz database\footnotemark[3], \footnotetext[3]{http://www.pmp.uni-hannover.de/cgi-bin/ssi/test/kurucz/sekur.html} and the Vienna Atomic Line Database.  We used IRAF ($\emph{splot}$) to fit the line profiles in the continuum-normalized spectrum with the line center, core depth, FWHM, and the local continuum value as free fit parameters.  The equivalent width (EW) and a fitting error were calculated in each profile fit.  All of the lines analyzed were best fit with a Lorentzian profile and the fit-derived EWs were checked for agreement with direct numerical integration.   
Three different measurements were performed for each line, varying the interval of nearby continuum data points from which the continuum level was determined.  The reported EW is the average of the 3 measured EW values, and the total EW uncertainty includes the RMS of the 3 fit values added in quadrature with the average fitting error and the 1\% error from continuum normalization ($\S$ 2.1).   Measurement uncertainties range from $\sim$ 3\% -- 50\%, with the peak of the distribution at $\sim$ 10\%, depending on the strength of the line and the ability to establish a well-defined continuum.  Our line identification and measurement results are listed in Table 2. From 99 high-Z absorption lines, the average measured (heliocentric) radial velocity of GD 40, which includes its gravitational redshift, is 10 km s$^{-1}$, with a standard deviation of 2 km s$^{-1}$ (Figure 9). 
The average high-Z line velocity agrees with the average helium line velocity (9 $\pm$ 4 km s$^{-1}$) implying a coincident physical position with helium in the gravitational well,  an additional confirmation of atmospheric location for the absorbers.  

Shipman et al.$\;$(1995) tentatively identified lines of neutral oxygen in the 1$\;$\AA$\;$ resolution UV HST (FOS) spectrum, noting a possible blend with silicon.  However, Friedrich et al. (1999) re-analyzed those spectra to derive abundances, and attributed the uncertain features entirely to Si {\small I} and Si {\small II}.  Our ground-based optical/near-IR detection of oxygen in GD 40 clears up the previous ambiguity regarding this element.  Oxygen is an important element for study since it is the most abundant constituent of rocky planetary material and is a component of water.

Nickle is not unambiguously detected in the HIRES data, but two of the stronger expected transition lines are possibly associated with noisy features (e.g. $\lambda$3465.6 in Figure 5).   Aluminum was also tentatively identified previously in the FOS spectrum (Shipman et al. 1995), and fit for abundance by Friedrich et al. (1999), although the latter authors regard its detection as unconvincing.

\section{ELEMENT ABUNDANCES}

The basic method of determining elemental abundances in a stellar atmosphere is to compare the observed spectrum with a synthetic spectrum derived from a model of the star's atmosphere.  The input physics of our models are described in Koester (2009b).  Absorption line data were taken from the Kurucz database, the Vienna Atomic Line Database, and Melendez \& Barbuy (2009).   Key input parameters are the effective temperature, $T_{eff}$, surface gravity,  $g$, and the atmospheric abundances of elements relative to the dominant constituent, in this case, helium.  At GD 40's temperature, along with a very weak hydrogen feature, the strong helium lines indicate a helium-mass-dominated atmosphere.

Voss et al. (2007) derived the effective temperatures and surface gravities, of a large sample of helium-line-dominated (DB) WDs from the broadening of helium lines in the SPY spectra.  For GD 40, they found $T_{eff}$ = 15,300 $\pm$ 500 K with log $g$ set to 8.0 (cooling age = 0.2 Gyr, mass = 0.59 $\Msun$).  In DB WDs with effective temperatures below $\sim$ 16,500 K, log $g$ cannot be determined unambiguously from the helium line broadening, so Voss et al. (2007) adopted the typical value, log $g$ = 8.0.  Since 97\% of DB WDs with effective temperatures between 16,500 and 19,000 K have log $g$ in the range 7.7 to 8.3 (Voss et al. 2007), we assume a $\pm$0.3 dex uncertainty in log $g$. 

Our synthetic spectrum reproduces the observed absorption spectrum  (see Figures 1 -- 8), and we compare the EWs of the model lines with those measured in the data.  The EW measurements are translated into abundance values relative to helium by interpolation between models with different relative abundances.  In a representative curve-of-growth plot of the Si {\small II} $\lambda$3862.6 line in GD 40 (see Figure 10), the EW versus abundance relation is well defined.  For each element, the adopted abundance value is the weighted average of the set of abundances from individual lines of that element.  We choose to weight the contribution from each line by the strength of its detection, i.e.$\;$using weights = SNR at line center (column 4 of Table 2) = [line core depth]/[RMS nearby continuum].   This gives nearly the same result as weighting by the inverse square of propagated measurement errors. 
For each set of high-Z element lines, the standard deviation of the mean (SDOM) of the set of abundances are in the range 0.03-0.09 dex, which always exceeds (by 50-200\%) the measurement error propagated from EW uncertainties.  This is expected since systematic errors enter in going from EW to abundance.  We take the SDOM as the ``formal'' error of the abundance average.
To be conservative in estimating our errors we include the difference between weighted and unweighted averages as an additional contribution to the uncertainty.  Finally, we varied the model parameters by their uncertainties, $T_{eff}$ (15,300 $\pm$ 500 K) and log $g$ (8.0 $\pm$ 0.3), one at a time, to calculate a per-element abundance error associated with the model temperature and gravity uncertainties, respectively.  In Table 3 we list the derived abundances and uncertainties, relative to helium, for the 9 detected contaminants and we place useful upper limits on Al and Ni, as well as 6 additional elements.   For the abundances relative to helium, the $T_{eff}$ uncertainty is usually the dominant source of error, as can be readily seen from Table 3.  

With a precision of $\sim$ 0.1 -- 0.2 dex, our work reduces the abundance uncertainties for previously identified elements, Mg, Si, and Ca, by at least a factor of 5.   For Mg, Si, Ca and Fe, our new abundances are all lower than the nominal values in F99W02 by $\sim$ 30-40 \%, which for Mg, Si, and Ca are within the errors (see Figure 11).  Differences in model temperatures used for this analysis (15,300 K) and F99W02 (15,150 K) are not the cause of the systematic abundance offset, since at lower $T_{eff}$ the implied abundances are even smaller.  
The HIRES detection of H$\alpha$ (EW = 0.23 \AA) agrees with the SPY data.  However, based on an H$\alpha$ EW measurement of 1.0 \AA$\;$(Greenstein \& Liebert, 1990), F99W02 obtained a hydrogen abundance nearly an order of magnitude higher than the HIRES and SPY results.  In retrospect, the feature assigned to H$\alpha$ (see Figure 1b of Greenstein \& Liebert, 1990) was noise.
From their possible UV detection of Al, Friedrich et al.$\;$(1999) suggest log[n(Al)/n(He)] = -7.15, just below our detection limit of log[n(Al)/n(He)] $<$ -7.1.   In our composition analysis we include the small, but relevant contributions from Ni and Al using the HIRES-derived upper limits on their abundances.

In principle, $T_{eff}$ can also be evaluated from the abundance agreement between different ionization states of an element, since we expect a consistent value for the total number of atoms of the element.  Both Mg and Fe display lines from two different ionic states, Mg {\small I}/Mg {\small II} and Fe {\small I}/Fe {\small II}.    The separated abundances for the different ionic states and varied model input temperatures are shown in Figures 12 and 13.   Although the mean deviations of the abundances are close to the differences between the average values, it is possible to distinguish between models with different temperatures.  For both Mg and Fe, the ionization balance is improved at the lower temperature limit, and degraded at the higher temperature limit.  Also, the ionized species always reflects a lower abundance than the neutral atoms, likewise suggesting the model temperature is too high. With our presentation of the uncertainties in Table 3, it is straightforward to recover all the element abundances, to within $\sim$0.05 dex, that result from the model with $T_{eff}$ = 14,800 K (if ever in the future this becomes established), from the Table 3 data by subtracting the $T_{eff}$ uncertainty from the abundances listed in column 2.  Similarly, the 15,800 K abundances can be re-calculated by raising all the column 2 abundances up by the $T_{eff}$ uncertainty. Such potential shifts in model temperature do not significantly alter our results of the relative elemental abundances in the parent body causing GD 40's atmospheric pollution.  This is because the abundances of polluting elements all move together, either down or up, with varying $T_{eff}$;  the pollutant-to-pollutant ratios are less sensitive to the model temperature than are the pollutant-to-helium ratios.

\section{RESULTS AND DISCUSSION}

How do these abundances help us understand extrasolar planetary systems?

\subsection{Bulk Composition}
First we note that  GD 40 has accreted at least the mass of a large asteroid.  With a helium-dominated convective zone (CVZ) that contains a fraction $q$ = log ($M_{cvz}/M$) = -5.86 (Koester 2009a) of the white dwarf's total mass of 0.59 $\Msun$, there is 1.7 $\times$ 10$^{27}$ g of helium in the CVZ.
From this and Table 3 abundances, we find a total mass of pollutant material presently detected in the CVZ of 3.6 $\times$ 10$^{22}$ g (Table 4).  For comparison, the mass of asteroid Juno is 3.0 $\times$ 10$^{22}$ g with a radius of 117 km (Konopliv et al. 2006), although Konopliv et al.$\;$(2006) note that Juno has a near metallic density, 4.43 $\pm$ 0.43 g cm$^{-3}$.  A 3.6 $\times$ 10$^{22}$ g asteroid of density 3.0 g cm$^{-3}$ has a radius of 142 km.

Second, we identify the material to be in a category of terrestrial-planetary composition.  Figure 14 depicts the bulk composition of GD 40's pollutants and a variety of solar system objects for comparison.  It is immediately apparent how 
GD 40's volatile-poor abundances are very different from primitive solar system abundances, and we find the closest agreement with an Earth-like composition.  94\% of Earth's mass comes from just 4 elements: O, Mg, Si, and Fe (Allegre et al. 1995).  Ca, Ni and Al contribute another $\sim$ 5\%, and all other elements are minor constituents.  Similarly, by bulk mass, the material polluting GD 40's atmosphere is 88\% O, Mg, Si, and Fe.  Ca contributes another 6\%.  Upper limits for the CVZ mass percentages of Al (2.5\%) and Ni (1.3\%) have been included in the totals.  

Jura (2006) and Farihi et al. (2009) have pointed out that, for a handful of polluted WDs with carbon measurements, including GD 40, the polluting abundance of carbon relative to iron is down by a factor of 6 or more from the solar ratio.  The deficiencies of carbon and hydrogen are consistent with rocky matter.  By similar arguments as apply for GD 362, which have been described in detail by Zuckerman et al. (2007), based on the quantity and constitution of pollutants, an interstellar or cometary origin for the high-Z elements is ``exceedingly unlikely''. 

The origin of the hydrogen is less certain because unlike the high-Z elements, it does not settle below the outer convective zone, which results in an accumulation of all the primordial and accreted hydrogen over the WD lifetime.  Both Voss et al. (2007) and Dufour et al. (2007) find a trend of increasing hydrogen mass with increasing age in helium-dominated WDs which argues strongly in favor of an accreted, rather than primordial, origin.    GD 40's hydrogen abundance is consistent with these populations of helium-dominated WDs (Figure 1 of Jura et al. 2009a).  
Since we do not have a constraint on its arrival epoch, it is possible that some, or maybe all, of the presently observed hydrogen is a result of previous accretion of other planetary matter (whose high-Z components have settled), or due to accretion from the interstellar medium, and may not even be associated with the most recently accreted material.   
Alternatively, some, or possibly all, of GD 40's hydrogen could be tied to the currently observed heavy element pollution.   At best we can place some upper limits.  One path by which hydrogen may survive processes that result in a refractory-rich environment is through H$_2$O.   Jura \& Xu (2009) propose that water from accreted minor planets may have a role in the hydrogen ``pollutions'' of helium-dominated WDs.   Our measured hydrogen abundance (at its upper error limit) imposes an upper limit to the amount of water or ice which could have been accreted (in total for GD 40's entire WD cooling time) at 3.6 $\times$ 10$^{21}$ g.  If all of this water was accreted along with the present high-Z material, then an upper limit to the amount of water in the bulk pollutant composition is 10\%.  We show below, in $\S$4.3, that due to settling of the heavier elements, the upper limit is likely to be much less than this.
In either case GD 40 has not accreted a very large amount of water, or hydrogen, along with its high-Z material.

GD 40's atmospheric pollution is consistent with having accreted rocky material.  However, the atmospheric abundances, n(Z)/n(He), need not be the same as the accreted abundances, and going from the measured abundances to the accreted ones requires some insight and modeling.

\subsection{Accretion -- Diffusion}

The relationship between observed and accreted abundances is governed by the interplay of accretion and diffusion.  Key inputs are the characteristics of the encounter with the reservoir, the rate that matter is falling into the star and the rates by which heavy elements are settling (diffusing) downward.
Fundamentals of accretion -- diffusion theory have been described in detail by Dupuis et al (1993).  More recently, Koester \& Wilken (2006) and Koester (2009a) have quantified mixing zone mass fractions and settling times of commonly accreted elements for a variety of white dwarf atmospheres in the temperature range of interest.  Table 4 lists settling times, $t_{set}(Z)$ ($\equiv$ diffusion time, $\tau(el)$ in Koester 2009a), calculated specifically for GD 40's CVZ.

Following the models of Jura et al (2009a) and Koester (2009a), accretion is occuring from the observed disk of material and may be represented by an exponentially decaying mass function with a characteristic lifetime, $t_{disk}$ ($\equiv \tau_{acc}$ in Koester 2009a).  
We are observing the system at a time $t_{obs}$, which we would like to constrain, with respect to $t_{set}$ and $t_{disk}$.  While exact relations governing the abundance ratios can be found from equation 8 of Koester (2009a) and equation 8 of Jura et al. (2009a), here we summarize the characteristics of three limiting cases:  (1) In the early phase of accretion, $t_{obs} \ll t_{set}$, the accreted abundance ratios are directly the observed atmospheric ones.  (2) For times on the order of the diffusion time scales a steady state is approached, and if the reservoir is long-lived, $t_{set}\;\lesssim\;t_{obs}\;\ll\;t_{disk}$,  the steady state values of the accreted abundances are modified from the measured abundances by the ratios of the settling times (equation 7 of Koester 2009a, equivalently equation 16 of Jura et al. 2009a):

\begin{equation}
\left({n(Z_1) \over n(Z_2)}\right)_{accreted} =  {t_{set}(Z_2) \over t_{set}(Z_1)} \left({n(Z_1) \over n(Z_2)}\right)_{CVZ}
\end{equation}

The ratios of settling times are at most  a factor of 2.6 between detected elements studied in this work.
(3) After the accretion reservoir is spent,  $t_{disk} < t_{obs}$, the atmospheric element ratios deviate exponentially with time from the accreted ratios, as gravitational settling causes the heavier elements to sink more rapidly than the lighter ones.  This case leads to a significant departure from the accreted abundances.

Referring to GD 40 CVZ abundances shown in Figure 15, the high-Z ratios do not differ from the Earth or CI chondrite ratios by more than a factor of 4, (except for carbon), mostly the factor is 1-2 as is the case for O, Mg, Ti, Cr, Mn and Fe.  This is not at all the exponential deviations between observed and accreted abundances that are expected from gravitational settling if accretion has ended, paused, or significantly decreased.  Also, there is no trend of increasing depletion of element abundances (relative to Earth or chondrite abundances) with decreasing element settling times.  Thus, the relative atmospheric abundances are consistent with accretion being presently ongoing, which is further supported by the observational existence of a disk. 
This implies that we are either observing GD 40 in a steady state, at an early phase of accretion, or somewhere in between.   
Figures 14 and 15 depict these two limiting extremes of the accreted abundances. 

\subsection{Oxides}
The Earth-like composition of GD 40's pollutants is almost certainly not coincidental, but rather, it is probably linked to the stoichiometric balance of rock-forming oxides.  In the solar system rocky planetary bodies are primarily composed of mineral oxides (MgO, Al$_2$O$_3$, SiO$_2$, CaO, TiO$_2$, Cr$_2$O$_3$, MnO, FeO, Fe$_2$O$_3$, NiO, and minor constituents) and possibly metals, e.g. Fe, Ni, Cr, etc, particularly if a core has formed. 

It is a remarkable result that the quantities of heavy elements in GD 40's atmosphere are consistent with a rocky composition wherein heavy elements are largely contained within oxides.  In other words, the observed oxygen is of sufficient quantity to form minerals with the other observed elements, as is evident from computing the number of oxygen atoms needed to form oxides, Z$_{p(Z)}$O$_{q(Z)}$, (measured relative to He):
  
\begin{equation}
{n(O)_{oxides} \over n(He)} = \sum\limits_Z {q(Z) \over p(Z)} {n(Z) \over n(He)}
\end{equation}

\noindent With Table 3 observed element (Z) abundances, which directly represent early-phase accretion as defined in $\S$4.2, Equation 2 adds up roughly to 1.5 $\times$ 10$^{-6}$, significantly below the measured oxygen abundance, n(O)/n(He) = 2.5 $\times$ 10$^{-6}$.  Additionally, as discussed in $\S$4.1, some of the oxygen may be associated with water/ice, since a portion, or possibly all, of the atmospheric hydrogen could be coeval with the current accretion episode.  Constrained by the relatively low hydrogen abundance, at most 0.5 $\times$ 10$^{-6}$ of the observed n(O)/n(He) can originate from water.

Now we have the question of accounting for all the excess oxygen, since the refractory-rich, volatile-poor element abundances imply that unbound, volatile oxygen is not a likely vehicle.  What happens in the steady state?  In order to calculate the steady state abundances from Equation 1, pollutant-to-pollutant ratios are needed.  We list early phase and steady state abundance ratios, relative to oxygen, in Table 5.
To evaluate the oxygen budget, let's look at the oxide balance in greater detail.  
As is suitable for a rocky parent body, we assume oxygen is mainly carried in oxides or water.  
   Removing the helium dependence of Equation 2 and working with the oxygen ratios, a balanced oxygen budget requires:
   
\begin{equation}
{1 \over 2}{n(H_{par}) \over n(O)} + \sum\limits_Z {q(Z) \over p(Z)} {n(Z) \over n(O)} = 1
\end{equation}

We use $H_{par}$ to represent the undetermined fraction of the observed hydrogen that is associated with water in the parent body causing the current accretion.  Another ambiguity comes from the fact that there are three principle states of iron in the Earth: metallic Fe (as in Earth's core), FeO, and Fe$_2$O$_3$, but in the white dwarf's atmosphere we are measuring the sum of Fe atoms from all states.  Since we do not have a clear determination of the partitioning of Fe in the GD 40 parent body, in Table 5 we show sums for 3 different distributions of Fe in order to better understand the range of effects on the oxide balance.  Iron is potentially one of the main carriers of oxygen, and in order to account for the observed O abundance, we find that the Fe can not be all metallic.  From the steady state oxide balance, we need at least 45\% of the iron atoms in FeO (at least 30\% if the more oxygen-rich species, Fe$_2$O$_3$, is the principle state) in order to carry the oxygen.  This constraint means that by total mass the accreted material was likely less than 28\% metallic iron.

Referring to the oxygen budget of Table 5, if one includes all of the hydrogen as water and assumes that all of the iron is in the more oxygenated form, Fe$_2$O$_3$ (no metallic Fe), then, even at the upper limit of the uncertainty, the fraction of oxygen accounted for by early accretion phase abundances gets close to, but does not quite add up to 1.  In other words, the parent body abundances that would be associated with early phase accretion have too much oxygen compared to what can be carried by high-Z elements and water.  
A more likely scenario is that accretion has been going on long enough for the system to be in a steady state with heavier elements settling more rapidly than the oxygen. 
Indeed, we see in Table 5, the steady state abundances are in 
balance with respect to oxides (i.e. the oxygen budget adds up to unity), and in this configuration, the problem of too much oxygen in the parent body disappears.  

We learn from this exercise that the oxide balance favors the steady state accretion situation (represented by the steady state bulk composition in Figure 14 and the steady state abundances in Figure 15, with element-to-element ratios derivable from the steady state column of Table 5).
In order for the system to achieve a steady state, the disk lifetime, must be longer than the settling time scales $t_{disk} \gg t_{set} \sim 6 \times 10^5$ yr.   
By these arguments, a lower bound on GD 40's disk lifetime is $t_{disk} > 10^6$ years.  With GD 40 in, or near, an accretion -- diffusion steady state, then after just a few settling times, the total accreted mass would be $\sim$ a few $\times$ 10$^{23}$g, or about the mass of Vesta, the second most massive asteroid in our solar system (Konopliv et al. 2006).  We have constrained $t_{set} < t_{obs} < t_{disk}$, but we do not know how many settling times have passed.  Also, since the disk is opaque (Jura et al. 2007a) we cannot estimate its mass, so we do not know how much more mass may be waiting to accrete from there.    

Returning to the question of water content in the parent body, since we have argued in favor of accretion taking place for at least a couple of settling times, then in the steady state, an upper limit to the percent of water by mass is 3\%, decreasing nearly linearly with $t_{obs} \ll t_{disk}$ (equation 6 of Jura et al. 2009a).   The water content of a planetary body can be a distinctive measure.  By mass, the Earth's mantle\footnotemark[4]\footnotetext[4]{The light element composition of Earth's core is uncertain and may include an appreciable amount of  
hydrogen and oxygen.}, Ceres and Callisto are 0.03 -- 0.1\%, 25\%, and 50\% water, respectively (van Thienen et al. 2007, McCord \& Sotin 2005, Canup \& Ward 2002, respectively).  Meteorites originating from the outer asteroid belt (2.5--4 AU) are up to 10\% water by mass, while those originating from the inner asteroid belt ($\sim$ 2 AU)  range from 0.05--0.1\% water by mass (Morbidelli et al. 2000).

\subsection{An Extrasolar Minor Planet}

GD 40 has a disk within its tidal radius, circumstellar silicate dust emission, the mass of a large asteroid in its convective zone, and pollutant abundances consistent with rocky material.  The most plausible origin of the atmospheric heavy elements is a tidally disrupted planetary parent body, such as an asteroid/minor planet.  Presumably, this object was part of an ancient planetary system, had its orbit perturbed (Debes \& Sigurdsson 2002), such that it journeyed within the star's tidal radius and was disrupted (Jura 2003), forming a disk of material that is accreting onto the star.

In Figure 15 the element abundances are shown normalized to the primitive material CI chondrites to assist our interpretation of how the GD 40 parent body may have evolved from a primitive constitution.   We see an overall trend of increasing abundances with increasing condensation temperature indicative of thermal processing.  The close agreement with bulk Earth values for O, Mg, Cr, Mn, and Fe anchors the Earth-like abundance nature of the polluting body, permitting us to interpret the deviations, most notably the depletion of silicon and the build up of refractory elements, in terms of evolutionary processes.

A distinctive result we find is n(Si)/n(Mg) = 0.30 $\pm$ 0.11 in significant contrast to the ratio near unity for the bulk Earth, the Sun, chondrites, and nearby stars.  This is not part of the pattern of the thermal trend, nor can it be caused by differential gravitational settling, since Si and Mg have nearly the same diffusion times (see also Figure 15).  Our result is consistent with the F99W02 value of n(Si)/n(Mg) = 0.3$^{+0.4}_{-0.2}$.
In this paper, we have reduced the uncertainties over previous work enough to recognize that the primordial elemental mix in this evolved planetary system has been altered within the parent body.

Figure 16 depicts the distinction from cosmic abundances.  The GD 40 star and proto-planetary-system was probably not born with a peculiar Si/Mg ratio.  Presently no solar system object has an observed Si/Mg abundance with such a low ratio.  Perhaps the closest comparison to a measured material in our own solar system comes from the Pallasite meteorites, which are thought to derive from the mantle-core boundary of a differentiated body (i.e. physically stratified into mantle, core, and/or crust), and measure n(Si)/n(Mg) = 0.6 (Buseck 1977).  One possibility is that the distinctive Si/Mg ratio may be associated with mantle/crust differentiation, in which Si was concentrated in a crust, and Mg in a mantle, and the crust was later lost.  Evidence for eroding collisions exists in our solar system in the form of the Moon and the diversity of the population of asteroids and meteorites such as the Pallasites.  The so-called hit-and-run process (Asphaug et al. 2006), thought to play a role in terrestrial planet formation, results in increasingly exotic characteristics for the unaccreted remnants of planetesimal collisions.  If differentiation takes place in a planetary body, Si-rich minerals tend to rise into the crust, whereas Mg-rich minerals prefer the mantle.  In Earth's crust n(Si)/n(Mg) = 11, but in a large rocky planetary body such as Earth, the 35 km crust is a relatively thin layer and the mantle/crust mass ratio is 169:1.  Thus, even though Si is highly concentrated in the crust, a relatively small amount by mass has been depleted from the Earth's mantle.  In an asteroid sized body, such as Vesta (radius = 270 km), the crust is thought to be $\sim$27 km, and a mantle/crust mass ratio has been estimated at 3:1 (Ghosh \& McSween 1998, Jones 1984), making a substantial Mg and Si differentiation in an asteroid plausible.

A difficulty of this explanation is with the observed enhancements of Ca, Ti, and possibly Al in GD 40.  Similar to silicon, the elements Ca, Ti and Al also tend to concentrate in the crust of a differentiated body.  For instance, Si, Ca, Ti, and Al are respectively, 1.3, 1.8, 3.4 and 3.5 times more abundant in Earth's crust than in Earth's mantle (Palme \& O'Neill 2007).  If GD 40's parent body lost its crust, then should not Ca, Ti and Al also be depleted along with Si?  Something else must be going on.  Since Ca, Ti and Al are refractory, most likely their enhancements are attributable to a significant thermal process.
Indeed, Earth's mantle, while depleted in Si relative to primitive CI chondrites, is modestly enhanced in Ca, Ti and Al (see Table 6), notably due to the refractory lithophile nature of these elements.    Given the strong enhancements of Ca and Ti in GD 40's polluting planetary body, it likely experienced a more extreme thermal history than Earth, as may have come about for example from intense heating during the red giant phase or through the energy of collisions.
Another consideration is that exposure to high temperatures during planetary formation and/or evolution could be important.  With an estimated WD mass of 0.59 $\Msun$ for GD 40, its main sequence progenitor was likely greater than 1.5 $\Msun$ (Weidemann 2000).  As previously suggested by Jura et al. (2009a), the circumstellar environments around stars more massive than our Sun may support conditions of more intense heating of the protoplanetary material, planetesimals and planets than in the solar nebula, which could explain the dramatically high abundances of calcium polluting the white dwarfs GD 362 and GD 40.  
One possibility for GD 40 is that both heating and crust loss have occurred, and that Ca and Ti are enhanced because the heating effect on their abundances dominates over the crust loss.  In that case, if the GD 40 parent body had not lost its crust, the Ca, Ti, and Al abundances would be even more enhanced than what are measured now.

Fegley \& Cameron (1987) describe another mechanism  -- a silicate vaporization process -- which removes substantial quantities of material (e.g. $\sim$ 80\% of Mercury's silicate) from the outer layers of a differentiated body due to prolonged heating.  Their calculations were designed to explain the unusually high density of planet Mercury and are successful at doing so by modeling the protoplanet's exposure to a high-temperature (2500 K -- 3500 K) phase of the primitive solar nebula.  More generally we obtain a sense of the constitutional evolution of a planetary body with an initially chondritic composition after experiencing a period of intense heating.   The models predict a unique resulting abundance pattern such that, relative to chondritic abundances, the mantle+crust becomes depleted in SiO$_2$, while MgO, CaO, TiO$_2$, and Al$_2$O$_3$ are enhanced. In particular, model \#4 (from Table 4 of Fegley \& Cameron 1987) yields Si/Mg, Ca/Mg, and Ti/Mg ratios in agreement with GD 40's measured values, as shown in Table 6.    We do not refer to the iron abundances from the mantle+crust results of Fegley \& Cameron (1987) since in their model most of the iron resides in a core.  This silicate vaporization model can simultaneously reproduce the Si depletion, and the Ca and Ti enhancements in GD 40's polluting parent body, however, the calculated high density planet has an iron-to-silicate mass ratio twice that of Earth, which is appropriate for planet Mercury, but is not what we observe in GD 40.   Although this model involves differentiation of the parent body and essentially an erosion of the outer layers - mechanisms which we think are important - it is highly idealized and its application to real systems is uncertain. 

Finally we note that a crust/mantle differentiated body is likely to have also formed a core composed primarily of Fe metal as in Earth's core.
As discussed in $\S$4.3, within the uncertainties of GD 40's steady state pollutant abundances, the oxide balance allows up to 70\% of the iron to originate from a metallic form ($<$ 55\% if the iron oxide is predominantly FeO), although a rocky composition with no metallic iron is also allowed by the oxygen budget (Table 5).  This translates to an upper limit on the size of a metallic core in the accreted parent body at $\sim$25\% of its total mass (assuming a Vesta-like core composition of 0.89 Fe by mass, as modeled by Ghosh \& McSween 1998).  For comparison, Earth's core is 32\% of its mass (Allegre et al. 1995), and estimates of core sizes in Mars, Vesta, and the Moon are 30\%, 10\% and 5\% of their total masses, respectively (Righter \& Drake, 1996).   

\section{CONCLUSIONS}

Our Keck/HIRES observations and abundance analysis make GD 40 the second dusty polluted white dwarf (GD 362 is the first) with a comprehensive set of measured element abundances and at least the mass of a large asteroid in its outer convective layer.  GD 40's atmosphere is polluted by material with a rocky chemical composition, approximately similar to bulk Earth which is dominated by O, Mg, Si, and Fe.   The inferred parent body was not a predominantly icy object, containing at most 10\%, but more likely $<$ 3\%, water by mass.  The balance of elements within oxides and water implies that the system is in, or near, an accretion -- diffusion steady state, which places a lower bound on GD 40's disk lifetime at 
$t_{disk} >  1\; Myr$.
These results, taken together with the infrared $\it{Spitzer}$ observations, can best be explained by the accretion of an orbitally-perturbed, tidally-disrupted, ancient, planetary parent body of at least the mass of Vesta (radius $\sim$ 270 km), carrying with it a history of this extrasolar system's formation and/or evolution.  The suite of element abundances reveals enhanced Ca and Ti, probably due to thermal processing, and a distinctively low Si/Mg ratio which may be an indication that differentiation has occurred in this planetary system, since a minor planet stripped of its crust could conceivably cause the silicon deficiency.  Within the uncertainties of the oxygen budget, the accreted body may have had a metallic core up to one-quarter of its total mass.  Because a solid planetary body has conveniently been ground down to its elemental constituents, this dusty polluted white dwarf provides us a powerful tool for measuring the bulk chemical composition of an extrasolar minor planet, at a level of detail and precision greater than what is possible for most terrestrial planetary bodies in our own solar system.  \\
\\

We thank the referee, P. Bergeron, for a prompt review and useful comments, B. Hansen for helpful discussions, and the Keck Observatory staff for their support.
This work is supported by grants from the NSF and NASA to UCLA. The data presented herein were obtained at the W.M. Keck Observatory, which is operated as a scientific partnership among the California Institute of Technology, the University of California and the National Aeronautics and Space Administration. The Observatory was made possible by the generous financial support of the W.M. Keck Foundation.  We recognize and acknowledge the very significant cultural role and reverence that the summit of Mauna Kea has always had within the indigenous Hawaiian community.  We are most fortunate to have the opportunity to conduct observations from this mountain.

\begin{table}[htdp]

\caption{Observations}
\begin{center}
\begin{tabular}{lcccc}
\hline 
\hline

UT Date  & Collimator & $\lambda$ range  & exposure   \\
		&		  & 		(\AA) 	   &	(seconds)  \\
\hline
2006 Sep 09 & Red & 5,690 -- 10,200 & 	2400  \\
2007 Nov 20 & Blue & 3,110 -- 5,990   & 2 x 2700  \\
2008 Nov 14 & Red & 4,500 -- 9,000   & 1800 \\
2008 Nov 15 & Red & 4,500 -- 9,000  & 2 x 1800 \\
\hline
\end{tabular}
\label{Data taken with the HIRES spectrometer on the Keck 1 telescope at Mauna Kea Observatory}
\end{center}
\end{table}
$\mathbf{Notes.}$ The continuum SNR is discussed in $\S$2.1

\newpage
\begin{center}
\begin{longtable}{ l l c c c @{ $\pm$ }c}
\caption{GD 40 Absorption Lines in HIRES Spectra} \\

\hline
\hline
Ion	& $\lambda$$^a$ &	v$^b$   & SNR$^c$  & \multicolumn{2}{c}{EW$^d$}   \\
	&	(\AA)		   & (km s$^{-1}$)  & 	(line	  &  \multicolumn{2}{c}{ (m\AA)} \\
	&				&		&  center)		\\

\hline
\endfirsthead

\multicolumn{6}{c}{Table 2 --- \emph{Continued}} \\
\hline
\hline
Ion	& $\lambda$$^a$ &	v$^b$   & SNR$^c$  & \multicolumn{2}{c}{EW$^d$}   \\
	&	(\AA)		   &	(km s$^{-1}$)  & 	(line	  &  \multicolumn{2}{c}{ (m\AA)} \\
	&				&			&  center)		\\
\hline
\endhead

\hline
\endlastfoot

H {\small I} & 6562.8 & ...  & 2 & 230 & 80 \\ 
O {\small I} & 7771.9 & 8.1 & 6 & 172 & 8 \\ 
O {\small I} & 7774.2 & 9.8 & 5 & 149 & 15 \\ 
O {\small I} & 7775.4 & 10.6 & 3 & 116 & 18 \\ 
O {\small I} & 8446.4 & 13.4 & 3 & 250 & 40 \\ 
Mg {\small I} & 3832.3 & 11.9 & 5 & 84 & 8 \\ 
Mg {\small I} & 3838.3 & 10.7 & 7 & 140 & 10 \\ 
Mg {\small I} & 5172.7 & 6.1 & 2 & 35 & 13 \\ 
Mg {\small I} & 5183.6 & 10.7 & 2 & 57 & 11 \\ 
Mg {\small II} & 4481.2 & ... & 14 & 470 & 30 \\ 
Mg {\small II} & 7877.1 & 6.0 & 1 & 60 & 20 \\ 
Mg {\small II} & 7896.4 & 7.9 & 3 & 160 & 20 \\ 
Si {\small II} & 3856.0 & 10.2 & 7 & 67 & 6 \\ 
Si {\small II} & 3862.6 & 11.8 & 5 & 45 & 6 \\ 
Si {\small II} & 4128.1 & 10.6 & 2 & 28 & 10 \\ 
Si {\small II} & 4130.9 & 10.0 & 5 & 42 & 6 \\ 
Si {\small II} & 5056.0 & 12.2 & 3 & 49 & 10 \\ 
Si {\small II} & 6347.1 & 9.8 & 8 & 122 & 7 \\ 
Si {\small II} & 6371.4 & 10.4 & 6 & 70 & 4 \\ 
Ca {\small II} & 3158.9 & 12.2 & 12 & 445 & 12 \\ 
Ca {\small II} & 3179.3 & 11.8 & 14 & 511 & 15 \\ 
Ca {\small II} & 3181.3 & 13.5 & 5 & 140 & 30 \\ 
Ca {\small II} & 3706.0 & 5.0 & 7 & 150 & 60 \\ 
Ca {\small II} & 3736.9 & 5.4 & 15 & 237 & 10 \\ 
Ca {\small II} & 3933.7 & 11.8 & 32 & 2500 & 200 \\ 
Ca {\small II} & 3968.5 & 12.5 & 25 & 1540 & 150 \\ 
Ca {\small II} & 8498.0 & 11.2 & 5 & 162 & 14 \\ 
Ca {\small II} & 8542.1 & 13.3 & 12 & 1040 & 60 \\ 
Ca {\small II} & 8662.1 & 14.6 & 8 & 560 & 30 \\ 
Ti {\small II} & 3217.1 & 9.3 & 2 & 15 & 5 \\ 
Ti {\small II} & 3234.5 & 8.0 & 9 & 92 & 5 \\ 
Ti {\small II} & 3236.6 & 9.6 & 7 & 59 & 5 \\ 
Ti {\small II} & 3239.0 & 9.0 & 7 & 44 & 4 \\ 
Ti {\small II} & 3242.0 & 7.1 & 5 & 46 & 5 \\ 
Ti {\small II} & 3248.6 & 9.2 & 4 & 40 & 11 \\ 
Ti {\small II} & 3252.9 & 9.9 & 3 & 17 & 6 \\ 
Ti {\small II} & 3254.2 & 12.5 & 3 & 28 & 4 \\ 
Ti {\small II} & 3261.6 & 10.1 & 5 & 34 & 4 \\ 
Ti {\small II} & 3278.3 & 9.3 & 2 & 24 & 9 \\ 
Ti {\small II} & 3278.9 & 12.7 & 2 & 17 & 7 \\ 
Ti {\small II} & 3287.7 & 6.9 & 3 & 15 & 3 \\ 
Ti {\small II} & 3322.9 & 10.8 & 4 & 34 & 6 \\ 
Ti {\small II} & 3329.5 & 8.1 & 3 & 24 & 4 \\ 
Ti {\small II} & 3332.1 & 9.4 & 3 & 16 & 5 \\ 
Ti {\small II} & 3335.2 & 14.4 & 3 & 27 & 9 \\ 
Ti {\small II} & 3341.9 & 7.0 & 7 & 53 & 5 \\ 
Ti {\small II} & 3349.0 & 8.5 & 8 & 58 & 5 \\ 
Ti {\small II} & 3349.4 & 9.7 & 14 & 105 & 5 \\ 
Ti {\small II} & 3361.2 & 8.8 & 12 & 85 & 4 \\ 
Ti {\small II} & 3372.8 & 8.8 & 10 & 77 & 4 \\ 
Ti {\small II} & 3380.3 & 9.5 & 4 & 21 & 3 \\ 
Ti {\small II} & 3383.8 & 8.5 & 8 & 75 & 5 \\ 
Ti {\small II} & 3387.8 & 6.3 & 3 & 22 & 3 \\ 
Ti {\small II} & 3504.9 & 11.4 & 4 & 15 & 3 \\ 
Ti {\small II} & 3685.2 & 8.2 & 9 & 48 & 3 \\ 
Ti {\small II} & 3759.3 & 9.6 & 9 & 40 & 4 \\ 
Ti {\small II} & 3761.3 & 8.3 & 7 & 35 & 4 \\ 
Ti {\small II} + Fe {\small II} & 3154.2$^e$ & 9.8 & 7 & 97 & 9 \\
Cr {\small II} & 3118.6 & 11.8 & 4 & 29 & 5 \\ 
Cr {\small II} & 3120.4 & 11.9 & 4 & 70 & 30 \\ 
Cr {\small II} & 3125.0 & 9.5 & 7 & 83 & 7 \\ 
Cr {\small II} & 3128.7 & 13.7 & 2 & 20 & 5 \\ 
Cr {\small II} & 3132.1 & 8.5 & 8 & 92 & 6 \\ 
Cr {\small II} & 3147.2 & 12.3 & 3 & 26 & 14 \\ 
Cr {\small II} & 3180.7 & 10.0 & 5 & 30 & 10 \\ 
Cr {\small II} & 3197.1 & 11.9 & 4 & 16 & 4 \\ 
Cr {\small II} & 3358.5 & 6.7 & 2 & 20 & 4 \\ 
Cr {\small II} & 3360.3 & 9.5 & 2 & 21 & 6 \\ 
Cr {\small II} & 3368.0 & 10.1 & 7 & 45 & 4 \\ 
Cr {\small II} & 3403.3 & 9.7 & 3 & 16 & 4 \\ 
Cr {\small II} & 3408.8 & 9.8 & 5 & 27 & 3 \\ 
Cr {\small II} & 3422.7 & 7.9 & 4 & 27 & 4 \\ 
Mn {\small II} & 3442.0 & 9.9 & 6 & 25 & 5 \\ 
Mn {\small II} & 3460.3 & 9.9 & 2 & 23 & 5 \\ 
Mn {\small II} & 3474.1 & 11.1 & 3 & 15 & 5 \\ 
Mn {\small II} & 3482.9 & 8.5 & 2 & 18 & 4 \\ 
Fe {\small I} & 3570.1 & 5.3 & 2 & 46 & 11 \\ 
Fe {\small I} & 3581.2 & 6.2 & 3 & 26 & 4 \\ 
Fe {\small I} & 3749.5 & 8.3 & 3 & 15 & 3 \\ 
Fe {\small II} & 3114.3 & 7.0 & 3 & 28 & 7 \\ 
Fe {\small II} & 3167.9 & 8.6 & 4 & 31 & 6 \\ 
Fe {\small II} & 3170.3 & 4.9 & 2 & 18 & 2 \\ 
Fe {\small II} & 3177.5 & 10.5 & 5 & 30 & 5 \\ 
Fe {\small II} & 3183.1 & 11.1 & 2 & 22 & 2 \\ 
Fe {\small II} & 3186.7 & 12.0 & 3 & 18 & 7 \\ 
Fe {\small II} & 3192.9 & 9.2 & 3 & 19 & 4 \\ 
Fe {\small II} & 3193.8 & 11.6 & 6 & 41 & 3 \\ 
Fe {\small II} & 3196.1 & 11.0 & 5 & 37 & 4 \\ 
Fe {\small II} & 3210.4 & 10.6 & 6 & 54 & 5 \\ 
Fe {\small II} & 3213.3 & 11.6 & 13 & 105 & 6 \\ 
Fe {\small II} & 3227.7 & 10.8 & 14 & 141 & 4 \\ 
Fe {\small II} & 3247.2 & 9.5 & 3 & 33 & 11 \\ 
Fe {\small II} & 3255.9 & 6.7 & 3 & 28 & 5 \\ 
Fe {\small II} & 3258.8 & 8.5 & 3 & 32 & 7 \\ 
Fe {\small II} & 3259.1 & 10.7 & 4 & 28 & 4 \\ 
Fe {\small II} & 3493.5 & 11.2 & 3 & 17 & 6 \\ 
Fe {\small II} & 4233.2 & 8.2 & 2 & 50 & 10 \\ 
Fe {\small II} & 4923.9 & 9.5 & 6 & 28 & 4 \\ 
Fe {\small II} & 5018.4 & 9.9 & 7 & 50 & 4 \\ 
Fe {\small II} & 5169.0 & 9.2 & 15 & 81 & 3 \\ 
Fe {\small II} & 5276.0 & 9.6 & 2 & 24 & 9 \\ 

\end{longtable}
  
\end{center}
$\mathbf{Notes.}$ -- Prominent He {\small I} lines are detected at wavelengths (in \AA): 3188, 3820, 3889, 3965, 4388, 4472, 4713, 4922, 5016, 5876, 6678 and 7281 (4026 and 7065 lie in gaps of the HIRES spectrum and were not accessible).  The average He line velocity is 9 km s$^{-1}$ with a standard deviation of 4 km s$^{-1}$. \\
$^a$ Laboratory wavelength in air. \\
$^b$ Velocities are in a heliocentric rest frame;  H$\alpha$ has an asymmetric profile and slightly ambiguous line center (see also Figure 7);  Mg {\small II} 4481 is a blended triplet - the line centers have not been individually determined here and the EW is the total for the blend. \\
$^c$ Line depth (from profile fit) divided by the RMS of the nearby continuum. \\
$^d$ $\S$2.2 describes the equivalent width measurements and uncertainties. \\
$^e$ The 3154.2 line is an unresolved blend of Ti {\small II} + Fe {\small II} (Figure 2), not used in the abundance determination. \\

\label{linelist}

\begin{table}[htdp]
\caption{GD 40 Atmospheric Element Abundances}

\begin{center}
\begin{tabular}{l l c c c c}
\hline
\hline
Z  & log[n(Z)/n(He)]          &  \multicolumn{4}{c}{ \% uncertainties }\\ 
\cline{3-6}
	&			               & formal &weighted & $T_{eff}$  &  log $g$ \\
\hline
H & -6.16 {\small  +0.14/-0.20}     & 	35	&   ... 		&	10	& 3	      \\
O  & -5.61  {\small +0.08/-0.09}    & 	9	&	2		&	17	& 1 		\\
Mg & -6.24 {\small  +0.12/-0.16}    &	13	&	6		&	25	& 11 	\\ 
Si & -6.76  {\small +0.07/-0.08}     &	7	&	1		&	8	& 13 	\\
Ca & -6.88 {\small +0.14/-0.22}     &	6	&	4		&	33	& 20	\\
Ti & -8.61 {\small +0.13/-0.20}      &	7	&	1		&	35	& 6	 	\\
 Cr & -8.31 {\small +0.12/-0.16}     & 	9	&	1		&	29	& 6		\\
Mn & -8.62 {\small +0.13/-0.18}     &	13	&	4		&	31	& 4		\\
Fe & -6.48 {\small +0.09/-0.12}     &	7	&	2		&	22	& 1		\\
\hline
	& previous UV \\
C & -7.2 {\small +0.2/-0.2} \\
\hline
 \hline
 &   upper limits  &   \multicolumn{3}{l}{from the absence of ($\lambda$/\AA)} \\
 \hline
N & $<$ -5.3 &  		\multicolumn{3}{l}{8680.3, 8216.3}  \\
Na & $<$ -7.0 &  	\multicolumn{3}{l}{5890.0, 5896.0}  \\
Al & $<$ -7.1 &  	\multicolumn{3}{l}{3961.5, 3944.0}  \\
P & $<$ -8.2 & 	 	\multicolumn{3}{l}{3509.0, 3786.6}  \\
Sc & $<$ -9.8 & 	\multicolumn{3}{l}{3572.5, 3613.8}	\\
V & $<$ -9.0 &	 	\multicolumn{3}{l}{3118.4, 3125.3}	\\
Ni & $\leq$ -7.7 & 	\multicolumn{3}{l}{3407.3, 3465.6$^\dag$, 3514.0$^\dag$}	\\ 
Sr & $<$ -10.1 &  	\multicolumn{3}{l}{4077.7, 4215.5}	\\
\hline

\end{tabular}
\end{center}
$\mathbf{Notes.}$ -- Abundances derived assuming a model with $T_{eff}$=15,300K and log $g$=8.0.  The listed number abundance errors are the sum of contributing uncertainties added in quadrature.
 Carbon abundance from UV data (Friedrich et al. 1999, Wolff et al. 2002). \\
``Formal'' is the standard deviation of the mean for the set of element lines. For H with just one line the formal error is the measurement error. \\
``Weighted'' = half the difference between weighted and unweighted averages. \\
``$T_{eff}$'' and ``log $g$'' errors come from varying the model parameters by their uncertainties, $T_{eff}$=15,300$\pm$500 K and log $g$=8.0$\pm$0.3.  Since the element abundances all move either up or down together with varying temperature, the $T_{eff}$ uncertainty directly represents this systematic shift, revealing the abundances that result from models with 14,800K and 15,800K (see last paragraph of $\S$3).\\
$^\dag$marginal features may be present at these wavelengths (e.g. NiII 3465.6 in Figure 5)\\
\end{table}

\begin{table}[htdp]
\caption{Gravitational settling times, contaminant masses and steady state mass accretion rates for GD40's convective zone.}
\begin{center}
\begin{tabular}{c l c r @{.} l r @{.} l c}
\hline
\hline 
\\
atomic \# & Z  &  log ($t_{set}$/yrs)  &  \multicolumn{2}{r}{$M_{cvz}/(10^{21}$ g)} & \multicolumn{2}{r}{$\dot M_{acc}/(10^{8}$  g s$^{-1}$)}  \\ 
         &	    &			 \\
\hline
  1 &   H  &  floats	&  0 & 29 $\pm$ 0.11 & ...  \\
  6 &   C  &  5.7943 & 0 & 32 $\pm$ 0.19 & 0&16 $\pm$ 0.10 \\
  7 &   N  &  5.7720 & $<$ 29&7 & $<$ 15&9 \\
  8 &   O  &  5.7587 & 16&6 $\pm$ 3.2 & 9&2 $\pm$ 1.8\\
11 &  Na &  5.6680 & $<$ 1&0 & $<$ 0&7 \\
12 & Mg &   5.6623 & 5&9 $\pm$ 1.9 & 4&1 $\pm$ 1.3 \\ 
13 & Al   &  5.6625 & $<$ 0&91 & $<$ 0&63 \\
14 &  Si  &  5.6520 & 2&06 $\pm$ 0.35 & 1&46 $\pm$ 0.25  \\
15 &  P  &  5.5986 & $<$ 0&083 & $<$ 0&066 \\
20 & Ca &  5.4694 & 2&2 $\pm$ 0.9 & 2&4 $\pm$ 0.8 \\
21 & Sc &  5.4389 & $<$ 0&003 & $<$ 0&004\\
22 &  Ti &  5.4373 & 0&050 $\pm$ 0.018 & 0&058 $\pm$ 0.021\\
23 &  V &  5.4398 & $<$ 0&022 & $<$ 0&03 \\
24 & Cr &  5.4321 & 0&11 $\pm$ 0.03 & 0&13 $\pm$ 0.04 \\
25 & Mn & 5.4105 & 0&056 $\pm$ 0.019 & 0&069 $\pm$ 0.023 \\
26 & Fe &  5.4006 & 7&8 $\pm$ 1.9 & 9&8 $\pm$ 2.3 \\
28 & Ni &  5.3715 & $\leq$ 0&50 & $\leq$ 0&67 \\
38 & Sr &  5.2089 & $<$ 0&003 & $<$ 0&006  \\
\hline
\multicolumn{2}{l}{Total detected} & & 35&5 & 27&4\\
\hline
\\
\end{tabular}
\end{center}
$\mathbf{Notes.}$ -- Settling times, $t_{set}$ ($\equiv$ diffusion times, $\tau_{diff}$) are defined in Koester (2009a). \\
Contaminant masses are calculated from Table 3 abundances and a helium-dominated convection zone mass of $M_{cvz}$(He) = 1.7 $\times$ 10$^{27}$ g. \\
In the steady state limit, $\dot M_{acc}$(Z) = $M_{cvz}$(Z)/$t_{set}$(Z). \\
Some upper limits, nitrogen in particular, are not very restrictive but are included nonetheless for completeness. \\

\end{table}

\begin{table}[htdp]
\caption{Limits of Parent Body Abundances}
\begin{center}
\begin{tabular}{l l l l c }
\hline
\hline
Z		&	n(Z)/n(O)		&	n(Z)/n(O)	\\
		&	Early Phase	& Steady State & \\
\hline
H	& 0.28 $\pm$ 0.11		&  $<$ 0.2  \\
C	& 0.026 $\pm$ 0.016    & 0.024 $\pm$ 0.015	 \\
Mg	& 0.24	$\pm$ 0.06		&	0.29	$\pm$	0.08  \\
Al	& $<$ 0.032	 			&	$<$ 0.04   \\
Si	& 0.071	$\pm$ 0.016		&	0.09	$\pm$	0.02  \\
Ca	& 0.053	$\pm$ 0.015		&	0.10	$\pm$ 	0.03   \\
Ti	& 0.0010 	$\pm$ 0.0002		&	0.0021 $\pm$	0.0004  \\
Cr	& 0.0020	$\pm$ 0.0004		&	0.0042	$\pm$	0.0008  \\
Mn	& 0.0010	$\pm$ 0.0003		&	0.0022	 $\pm$	0.0006  \\
Fe	& 0.13		$\pm$ 0.02			&	0.30	$\pm$	0.05	\\
Ni	& $<$ 0.008				&	$<$ 0.02 \\
\hline

Oxygen budget: \\
H$_2$O & $<$ 0.2  &  $<$ 0.1  \\
oxides (3 cases):	\\
all Fe in Fe$_2$O$_3$ &	0.69 $\pm$ 0.09	&	1.12 $\pm$ 0.12 \\
all Fe in FeO			&	0.62 $\pm$ 0.09	&	0.97 $\pm$ 0.10 \\
${1 \over 2}$ FeO, ${1 \over 2}$ Fe metal &	0.56 $\pm$ 0.09 & 0.82 $\pm$ 0.09 \\

\hline

\end{tabular}
\end{center}
\label{default}

$\mathbf{Notes.}$ -- Upper portion:  two limiting extremes for the accreted abundance ratios relative to oxygen.  Early phase and steady state are defined in $\S$4.2.
The pollutant-to-pollutant ratios are less sensitive to variations in T$_{eff}$ than are the pollutant-to-helium ratios, which results in smaller errors (30-60\% smaller) listed here than what one obtains in propagated errors from Table 3.   With the exception of hydrogen, the steady state values have been calculated using the settling times from table 4 according to Equation 1.  For hydrogen, the steady state fraction relative to O depends on the number of settling times that have passed.  An upper limit is estimated assuming $t_{obs} \geq 2\;t_{set}$, i.e. n(H)/n(O) $<$ (0.28+0.11)/2 (= early phase upper limit / 2 settling times). \\
Lower portion: 
The oxygen budget, which includes Al and Ni at their upper limits, is the fraction of the observed oxygen that may be accounted for by the total sum of oxides (MgO, Al$_2$O$_3$, SiO$_2$, CaO, TiO$_2$, Cr$_2$O$_3$, MnO, Fe in Fe$_2$O$_3$, FeO or Fe metal, NiO) and water.  Three different compositions involving iron are shown.     Uncertainties from the element ratios are propagated in quadrature. The oxide, Na$_2$O can also contribute, but we do not have an abundance measurement, or meaningful upper limit, for Na.  In the silicate Earth, Na$_2$O $\approx$ Cr$_2$O$_3$ $\approx$ 0.4\% by mass, within the errors of our oxygen budget.  The expectation is that the contribution from oxides and water will add up to unity (Equation 3).  The steady state abundances account for all of the observed oxygen, in agreement with a mineral stoichiometry.  \\

\end{table}

\begin{table}[htdp]
\caption{Distinctive Abundance Ratios}
\begin{center}
\begin{tabular}{l c c c c c l l}
\hline
\hline 
 Ratio & Silicate  & GD 40 &	GD 40	& F \& C&F \& C \\
	     &	 Earth & CVZ &	Steady State	&	model 4 & model 3 \\
\hline
 \multicolumn{6}{c}{Number ratios relative to CI chondrite values} \\
 \hline
Si/Mg	& 0.83 & 0.31 $\pm$ 0.12 &	0.32 $\pm$ 0.13 & 0.28 & 0.43  \\
Ca/Mg	& 1.16 & 4.0 $\pm$ 2.1 &	6.2 $\pm$ 3.3 & 3.0 & 4.1 \\
Ti/Mg 	& 1.16 & 1.8 $\pm$ 0.9 &	3.1	$\pm$ 1.6 	& 2.6 & 4.2	\\
Al/Mg 	& 1.16 &	$\leq$ 1.7		& $\leq$ 1.7		& 1.4 & 2.0	\\

\hline		

\end{tabular}
\end{center}
\label{}
$\mathbf{Notes.}$ -- According to Fegley \& Cameron (1987), mantle+crust vaporization of a planetary body due to exposure to high temperatures can result in the depletion of Si and the enhancement of Ca, Ti, and Al in the parent body.  Fegley \& Cameron data are from their Table 4.  Their model 4 representing a nonideal magma heated to 3500 K predicts mantle+crust ratios in agreement with GD 40's pollutant abundances.  Model 3 for a nonideal magma heated to 3000 K is also near agreement with GD 40 abundances.  Data for the silicate Earth (Earth's mantle+crust) are from McDonough \& Sun (1995); CI chondrites from Lodders (2003). \\
\end{table}

\newpage

\begin{figure}[htbp]
\begin{center}
\plotone{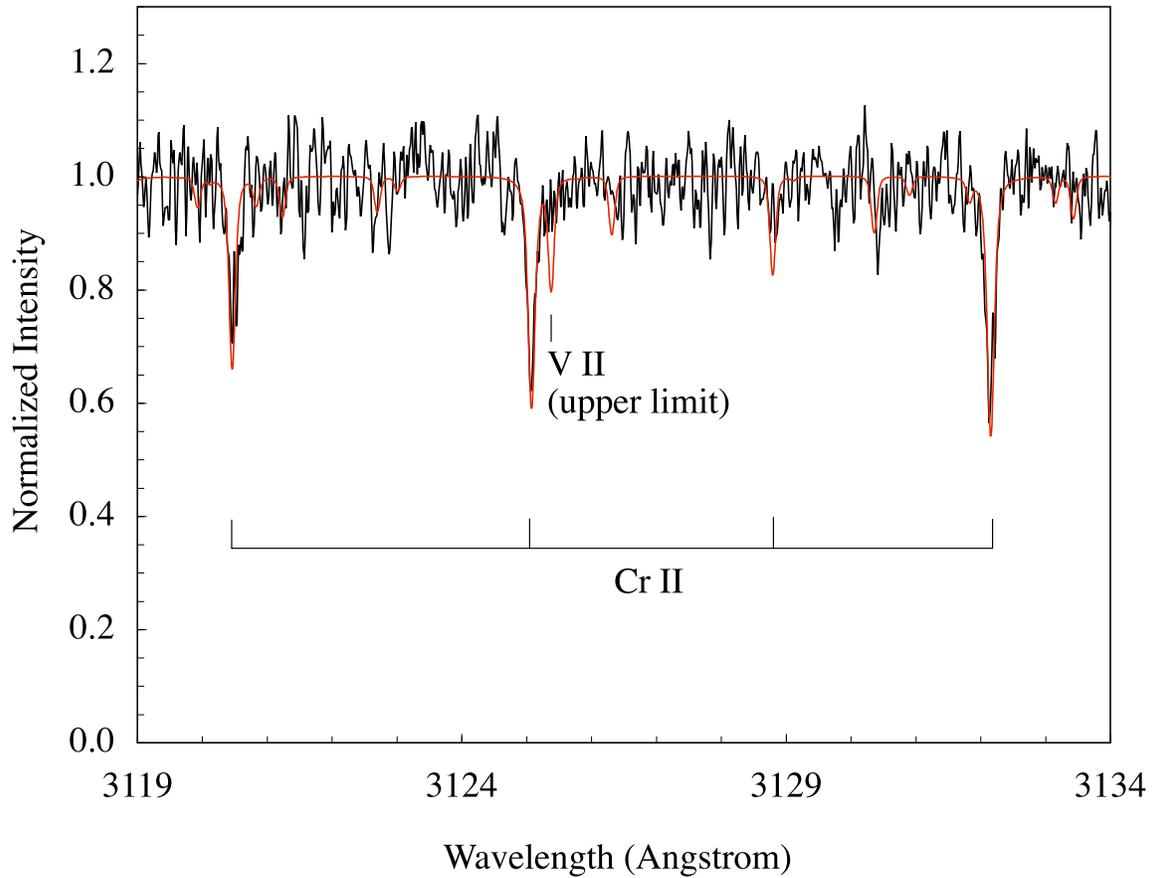}
\caption{A portion of the Keck/HIRES spectrum of GD 40.  Wavelengths are in air and a heliocentric rest frame.  The model spectrum shown in red has been convolved with a Gaussian profile to reflect the instrumental broadening (resolution, R $\sim$ 40,000), as measured by Gaussian fits to arc lamp lines.   The model is velocity shifted from the laboratory rest frame by the mean value, 10 km s$^{-1}$, measured from the data (see Figure 9).  The strongest Cr lines from Table 2 are displayed.}
\label{default}
\end{center}
\end{figure}

\begin{figure}[htbp]
\begin{center}
\plotone{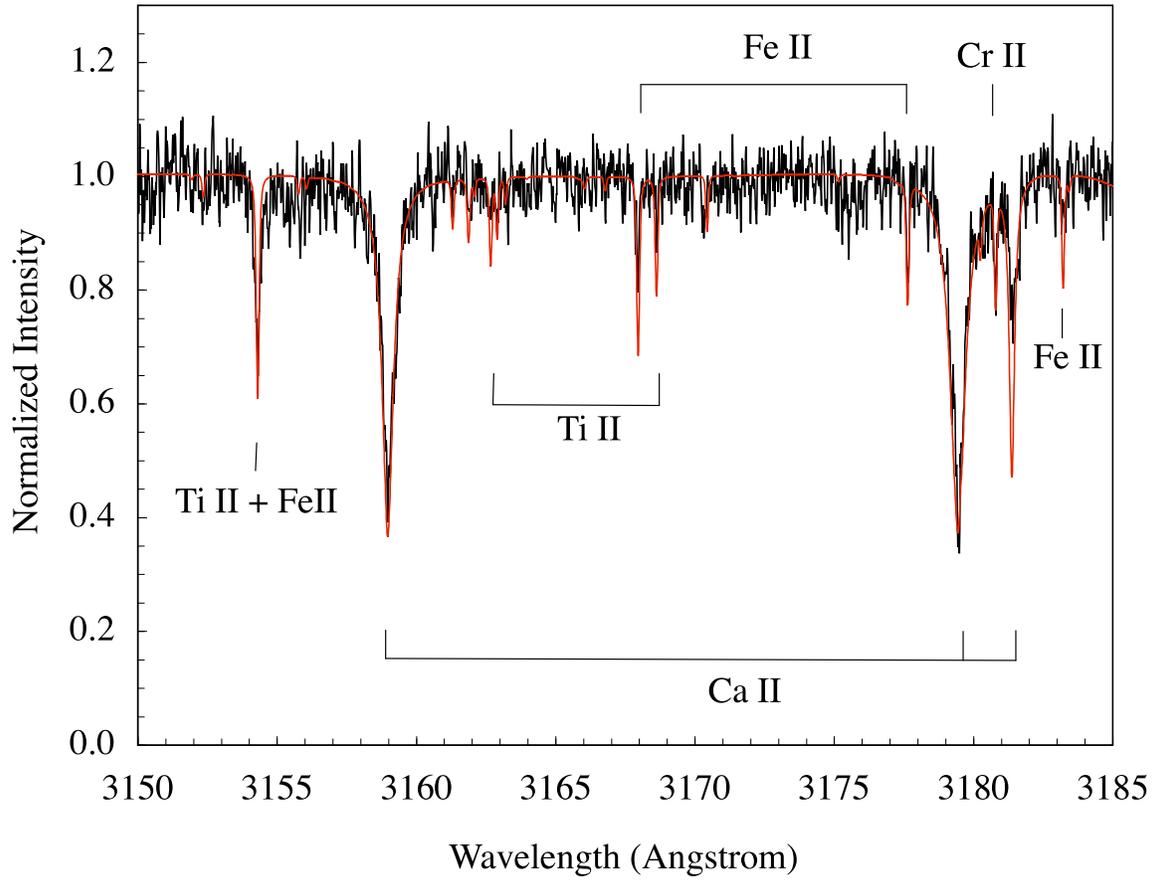}
\caption{Similar to Figure 1, with prominent Ca lines that do not originate from the ground state.}
\label{default}
\end{center}
\end{figure}

\begin{figure}[htbp]
\begin{center}
\plotone{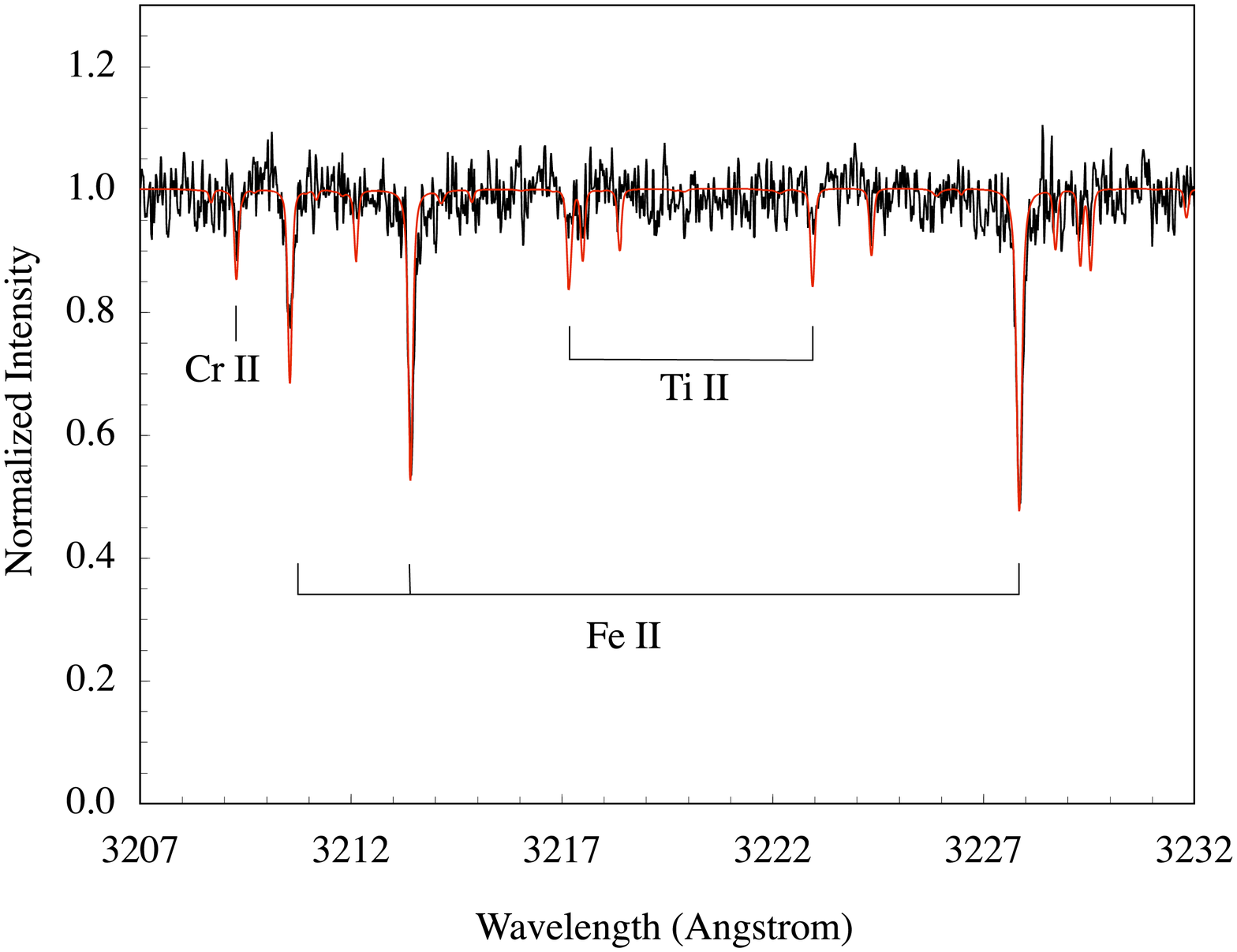}
\caption{Similar to Figure 1, with the strongest Fe lines from Table 2 displayed}
\label{default}
\end{center}
\end{figure}

\begin{figure}[htbp]
\begin{center}
\plotone{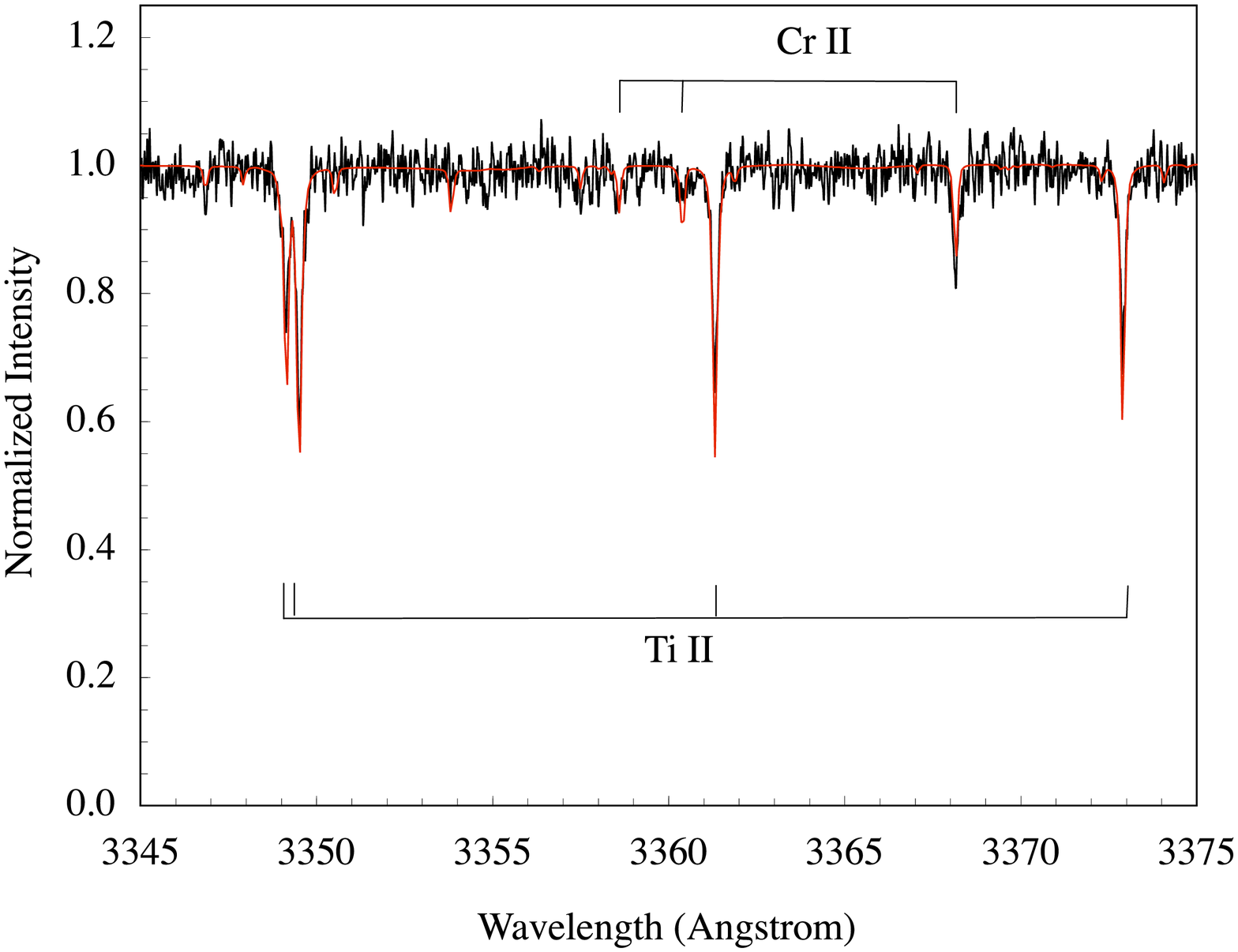}
\caption{Similar to Figure 1, with the strongest Ti lines from Table 2 displayed}
\label{default}
\end{center}
\end{figure}

\begin{figure}[htbp]
\begin{center}
\plotone{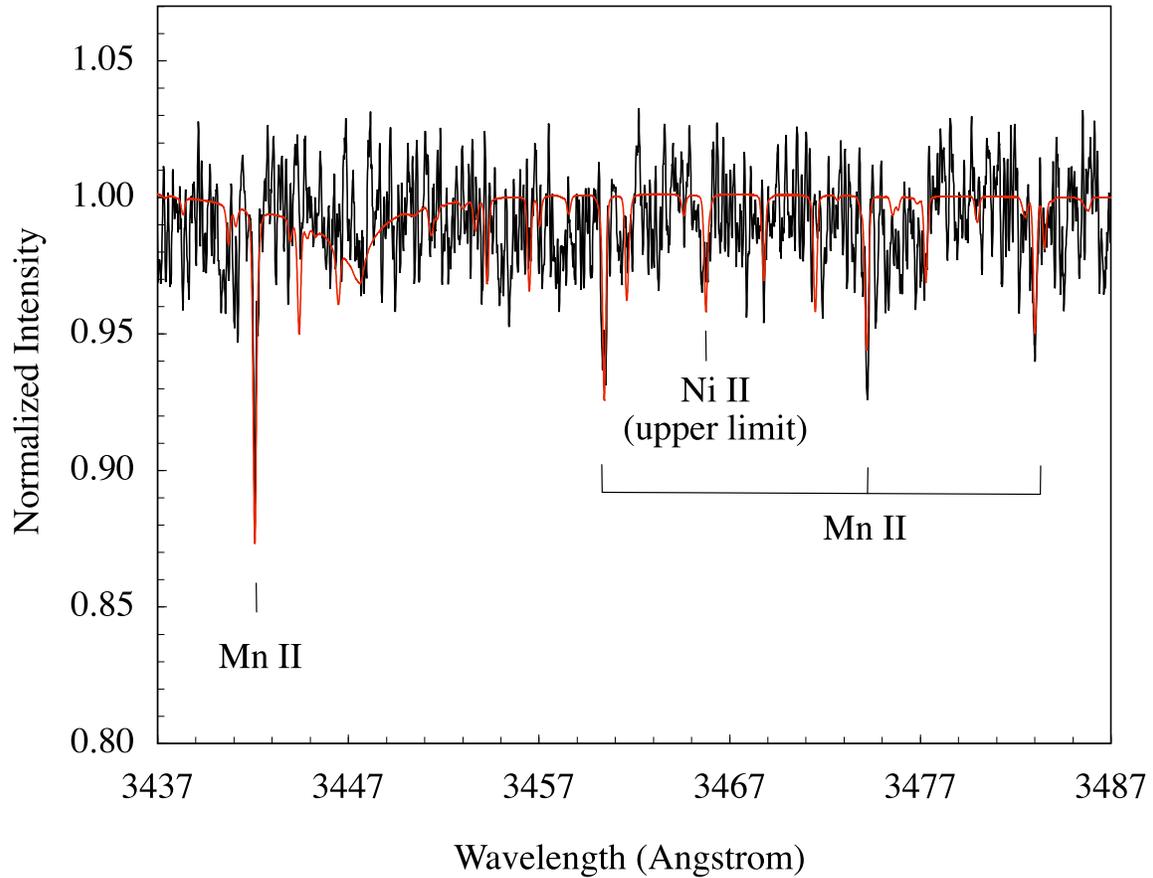}
\caption{Similar to Figure 1, but with all Mn lines from Table 2 displayed.  For ease of viewing all 4 Mn lines in one plot, these data are smoothed with a 7 point boxcar average, and the model resolution is matched to the smoothed data by convolution with a gaussian profile (which effectively corresponds to degrading the spectral resolution from R=40,000 to R=20,000).}
\label{default}
\end{center}
\end{figure}

\begin{figure}[htbp]
\begin{center}
\plotone{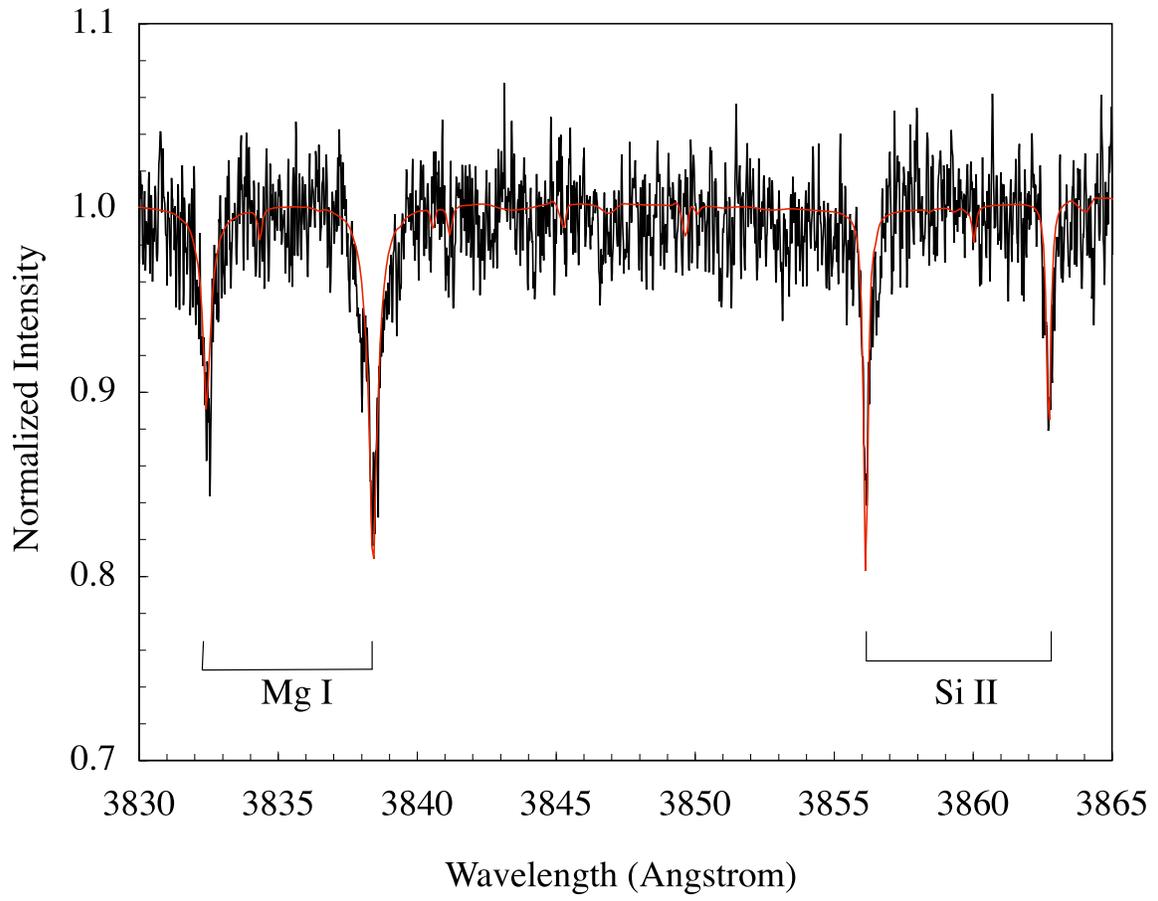}
\caption{Similar to Figure 1, but with prominent Mg and Si lines together.}
\label{default}
\end{center}
\end{figure}

\begin{figure}[htbp]
\begin{center}
\plotone{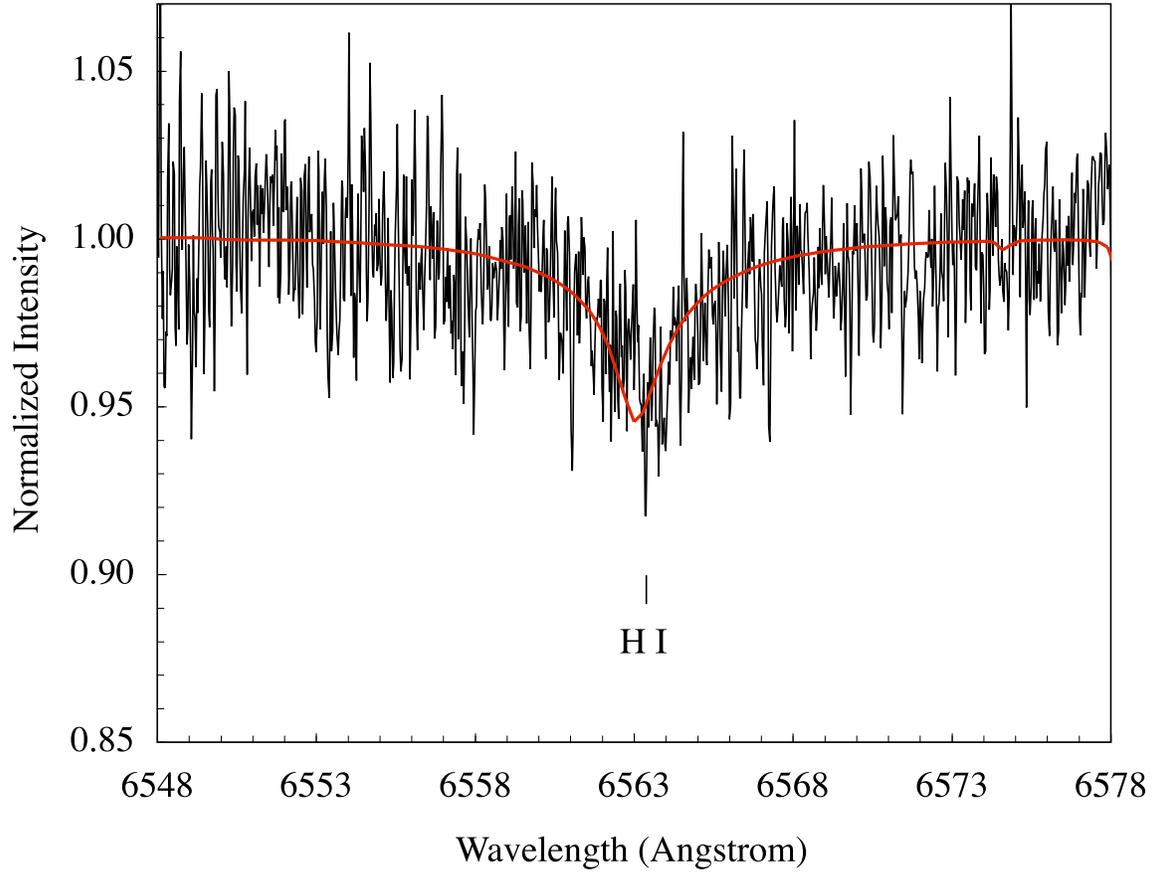}
\caption{Similar to Figure 1, but for H$\alpha$.  The line center in the data appears shifted from the systemic velocity (represented by the red model curve; discussed in $\S$2.2) by  $\sim$ +15 km s$^{-1}$.}
\label{default}
\end{center}
\end{figure}

\begin{figure}[htbp]
\begin{center}
\plotone{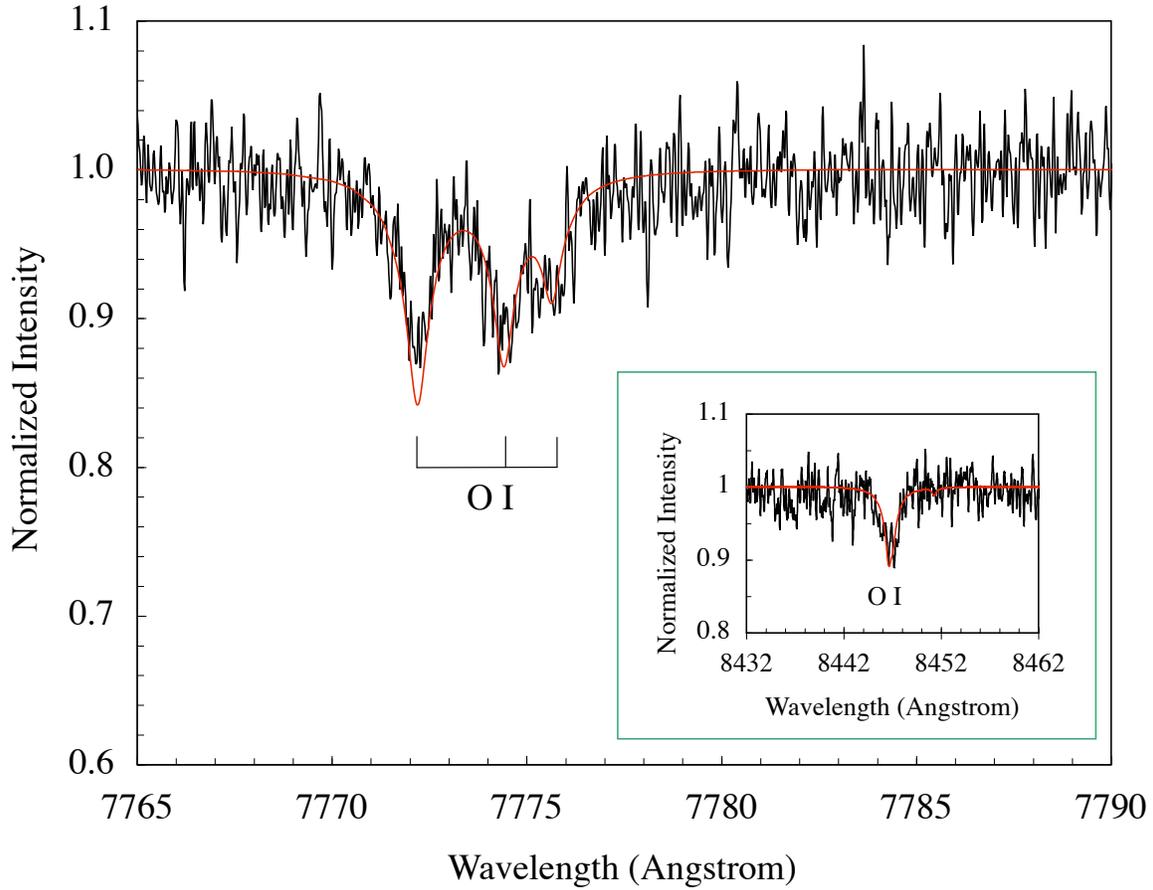}
\caption{Similar to Figure 1, but with all of the O lines from Table 2 displayed.  The inset data of the OI  $\lambda$8446 line are smoothed with an 11 point average.}
\label{default}
\end{center}
\end{figure}

\begin{figure}[htbp]
\begin{center}
\plotone{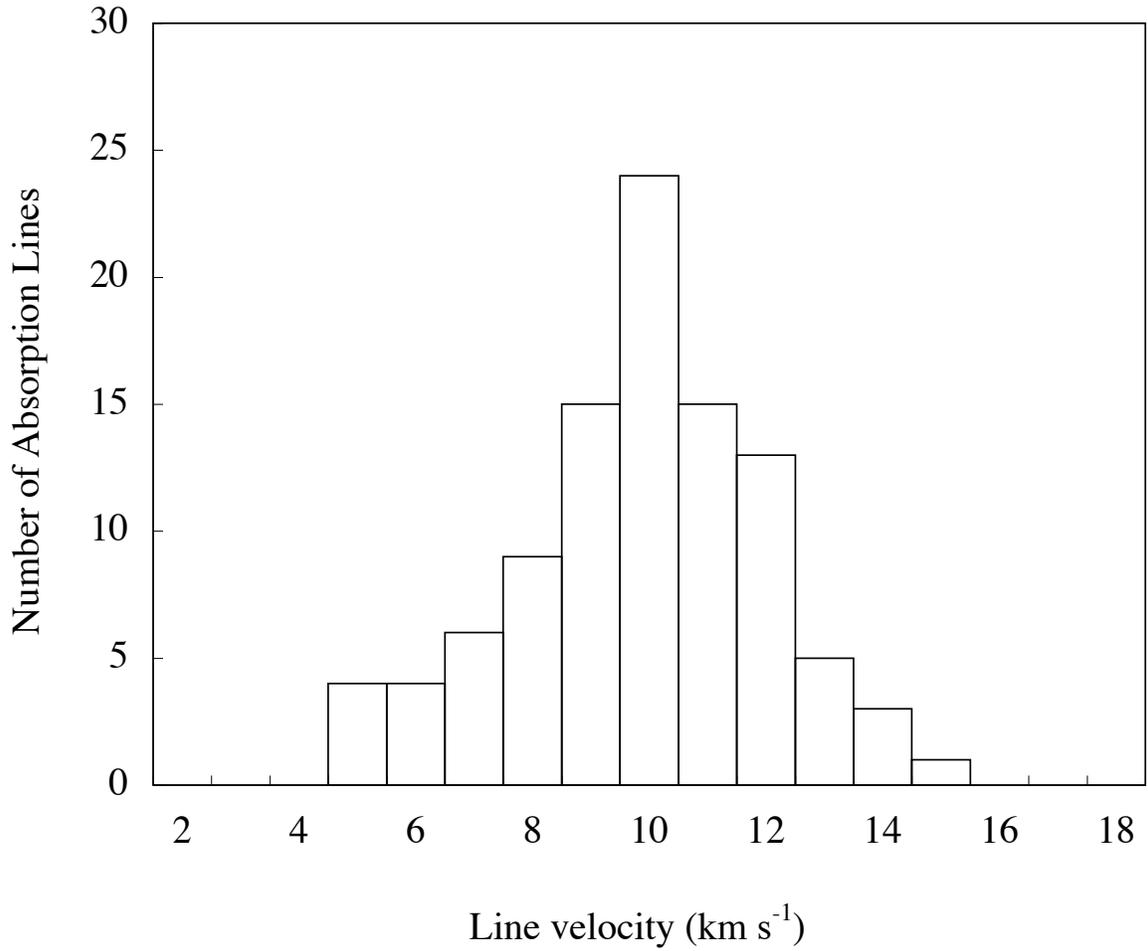}
\caption{Line velocities from Table 2.  All heavy element spectral lines in GD 40 have unambiguous identifications and their measured line centers lie within a distribution of velocities centered on a mean heliocentric value of 10 km s$^{-1}$. }
\label{default}
\end{center}
\end{figure}

\begin{figure}[htbp]
\begin{center}
\plotone{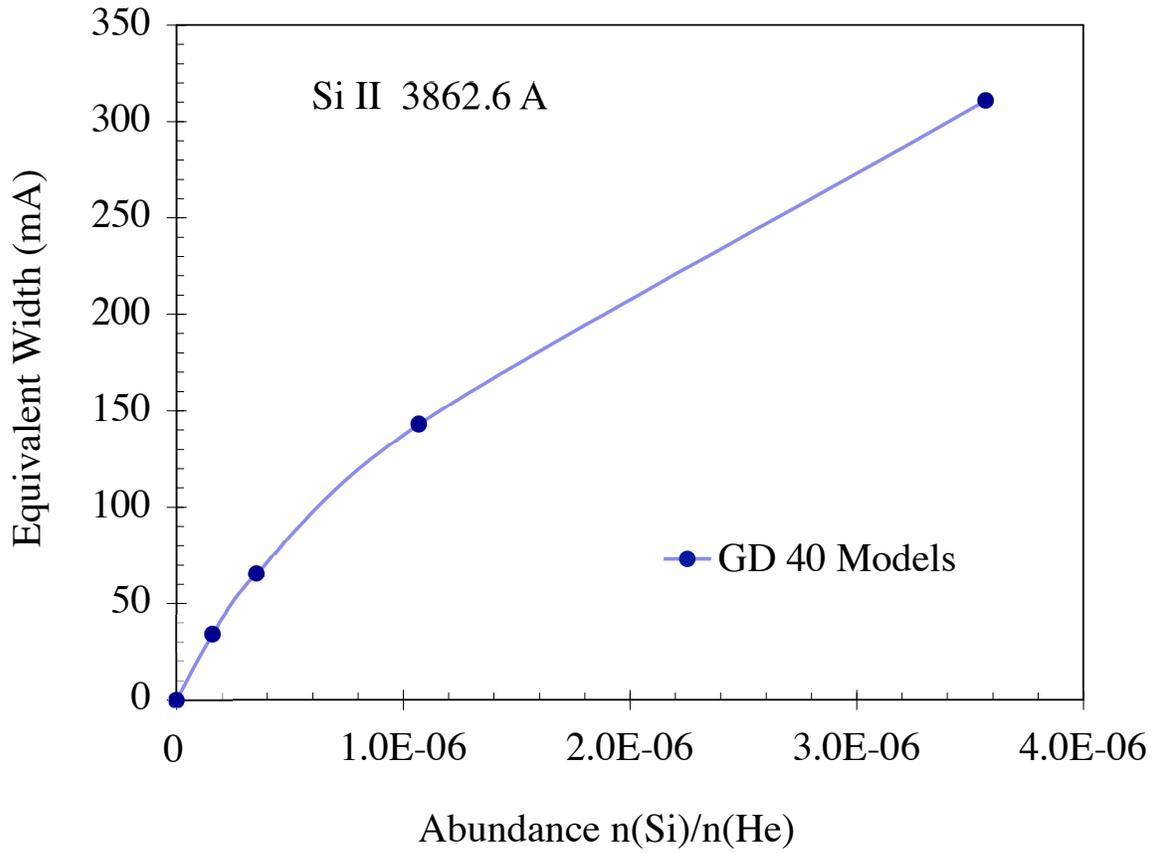}
\caption{Example curve-of-growth for the SiII $\lambda$3862.6{\AA} line in GD 40's atmosphere.    
The solid curve simply connects the points with a smoothed line.  There is a high fidelity between the model equivalent width and input Si abundance.  For a given absorption line the element abundance is determined by interpolation between models.}
\label{default}
\end{center}
\end{figure}

\begin{figure}[htbp]
\begin{center}
\plotone{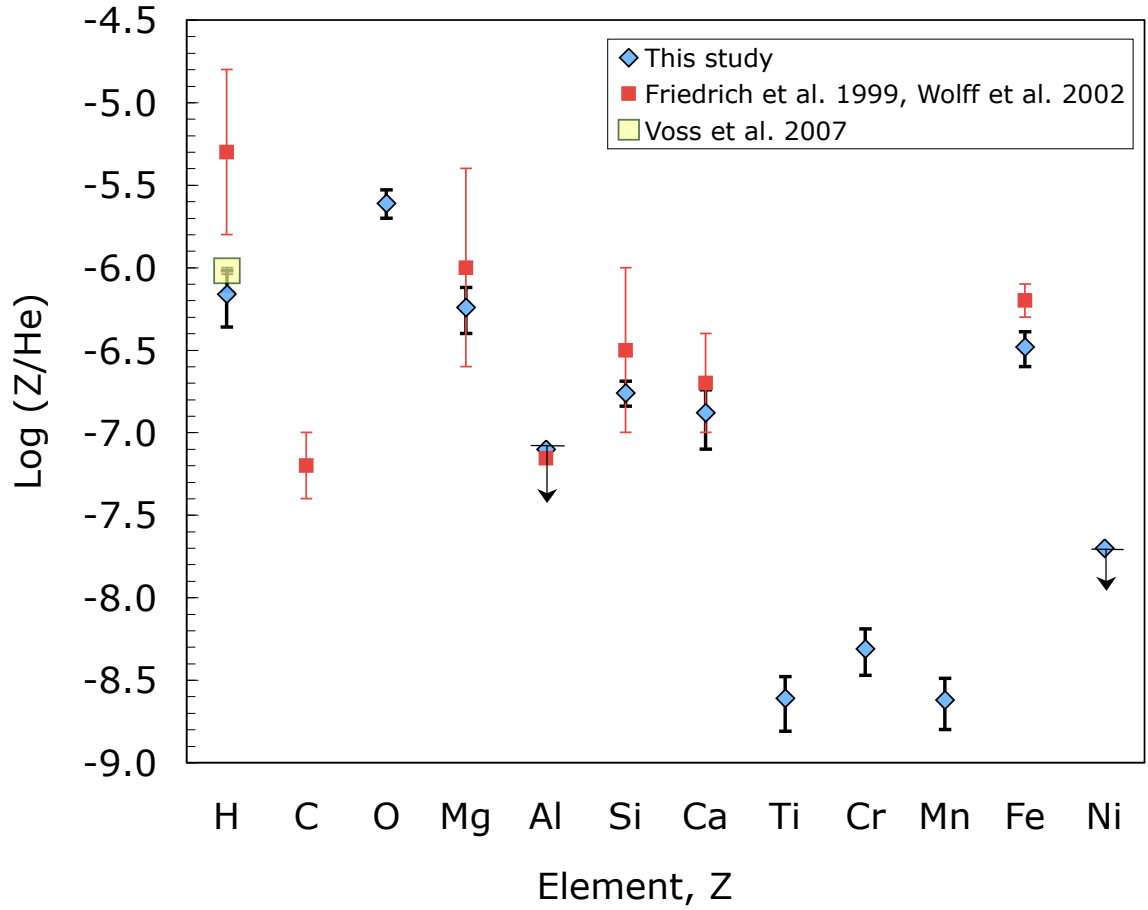}
\caption{GD 40's atmospheric abundances from Table 3 derived from Keck/HIRES optical spectra compared with previous abundances derived from UV data (C, Mg, Al, Si, Fe) and optical data (both previous H measurements).  Al and Ni are upper limits.}
\label{default}
\end{center}
\end{figure}

\begin{figure}[htbp]
\begin{center}
\plotone{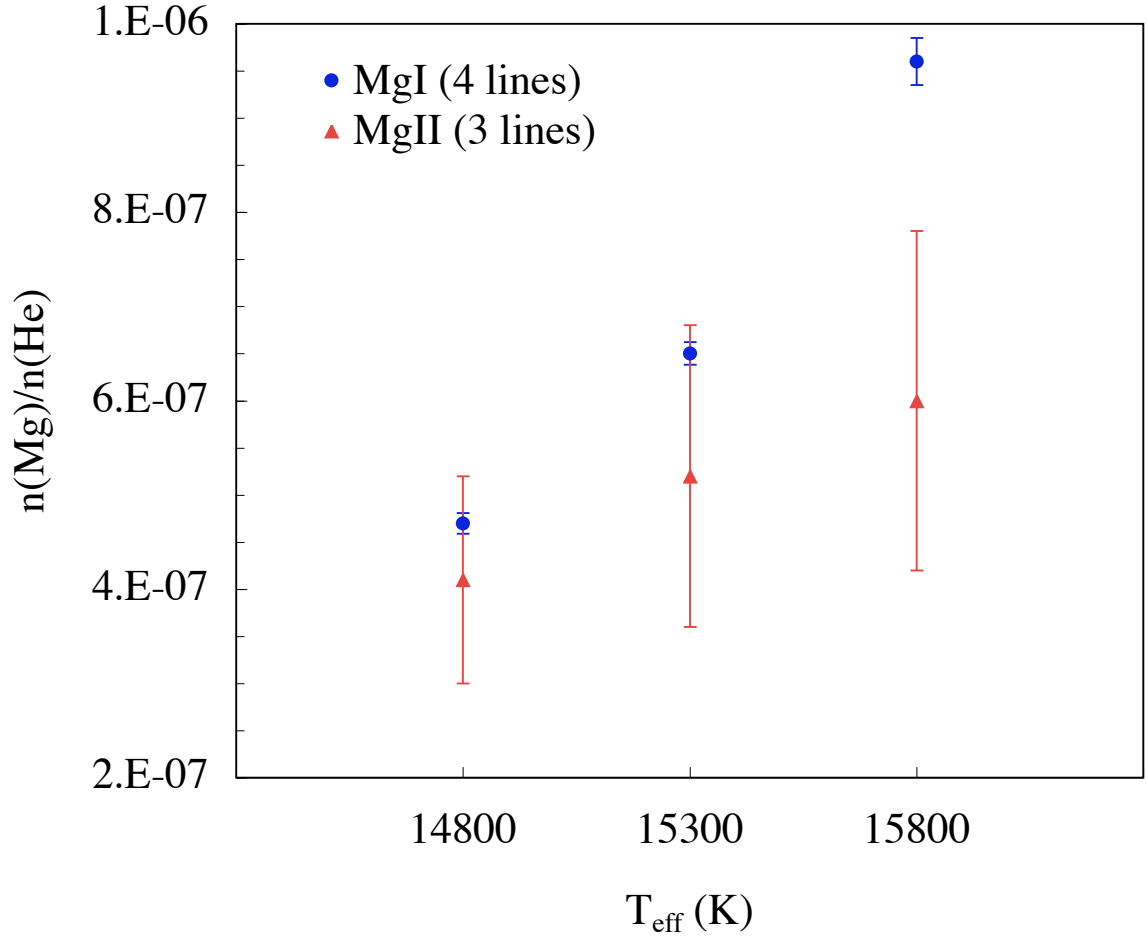}
\caption{Magnesium abundances derived from different ionic states of magnesium with models at low, central, and high values of GD 40's effective temperature.  Error bars represent one standard deviation of the mean added in quadrature with the difference between weighted and unweighted averages for the set of lines of the given ion.  The ionization balance is improved at the lower model temperature.}
\label{default}
\end{center}
\end{figure}

\begin{figure}[htbp]
\begin{center}
\plotone{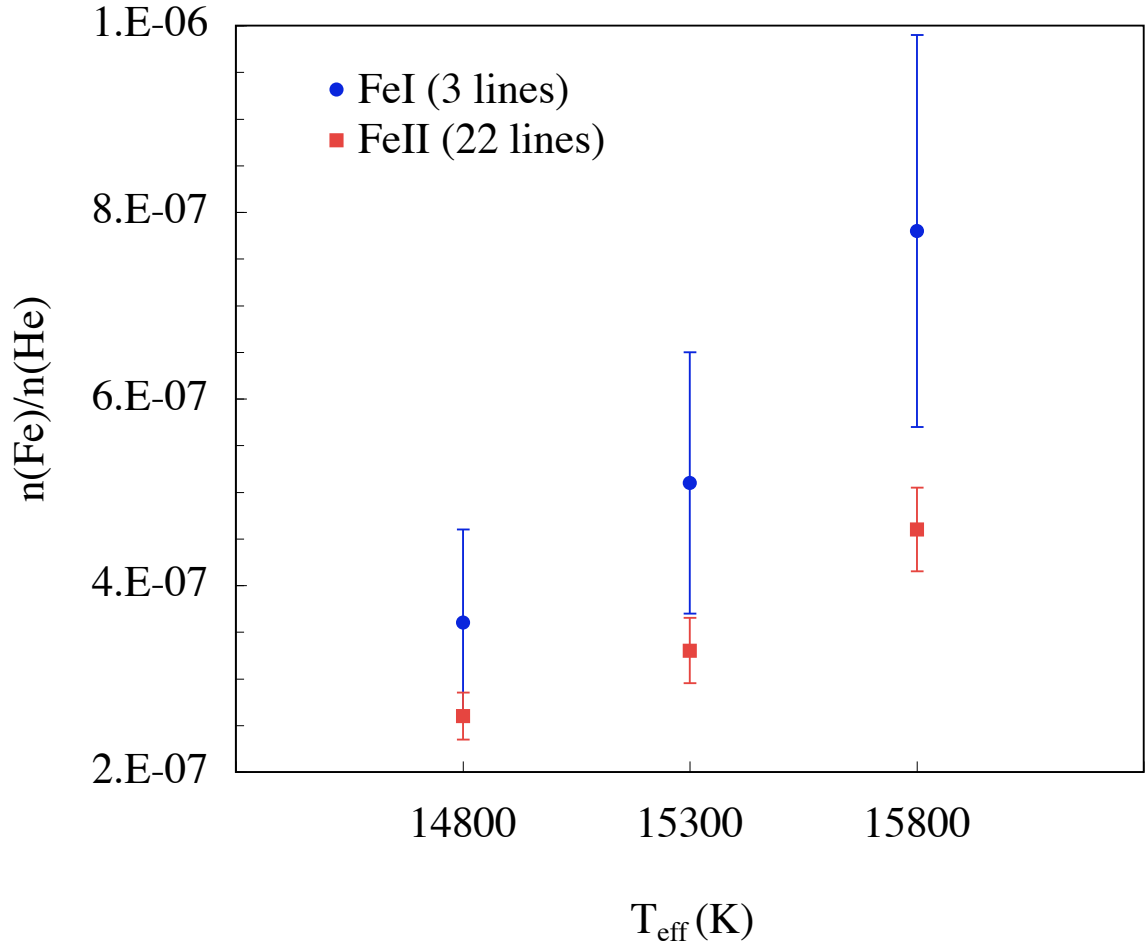}
\caption{Same as Figure 12, but for iron.}
\label{default}
\end{center}
\end{figure}

\begin{figure}[htbp]
\begin{center}
\plotone{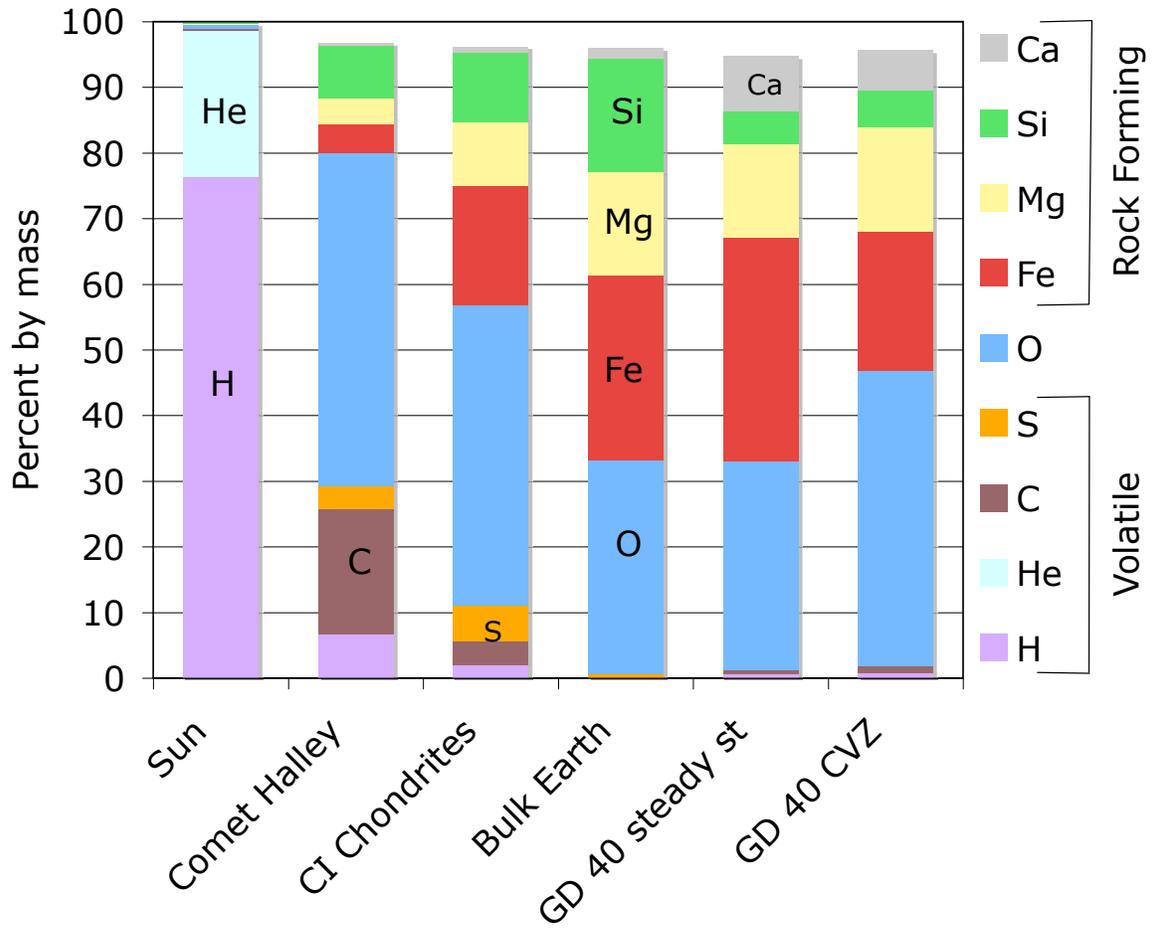}
\caption{Bulk compositions.  GD 40's volatile-deficient polluting abundances are most similar to a bulk terrestrial planet composition.  Totals do not reach 100\% because minor constituents are omitted.  GD 40's mass total includes Al and Ni at their upper limits (not shown).   Helium and sulfur are not measured in GD 40, but given their volatility, these elements are expected to be minor constituents.  GD 40 data are from Table 5 for the two limiting extremes of parent body abundances, steady state and CVZ (= convective zone $\equiv$ early phase); Sun and CI chondrites from Lodders (2003); comet Halley from Lodders \& Fegley (1998); bulk Earth from Allegre et al. (1995)}
\label{default}
\end{center}
\end{figure}

\begin{figure}[htbp]
\begin{center}
\plotone{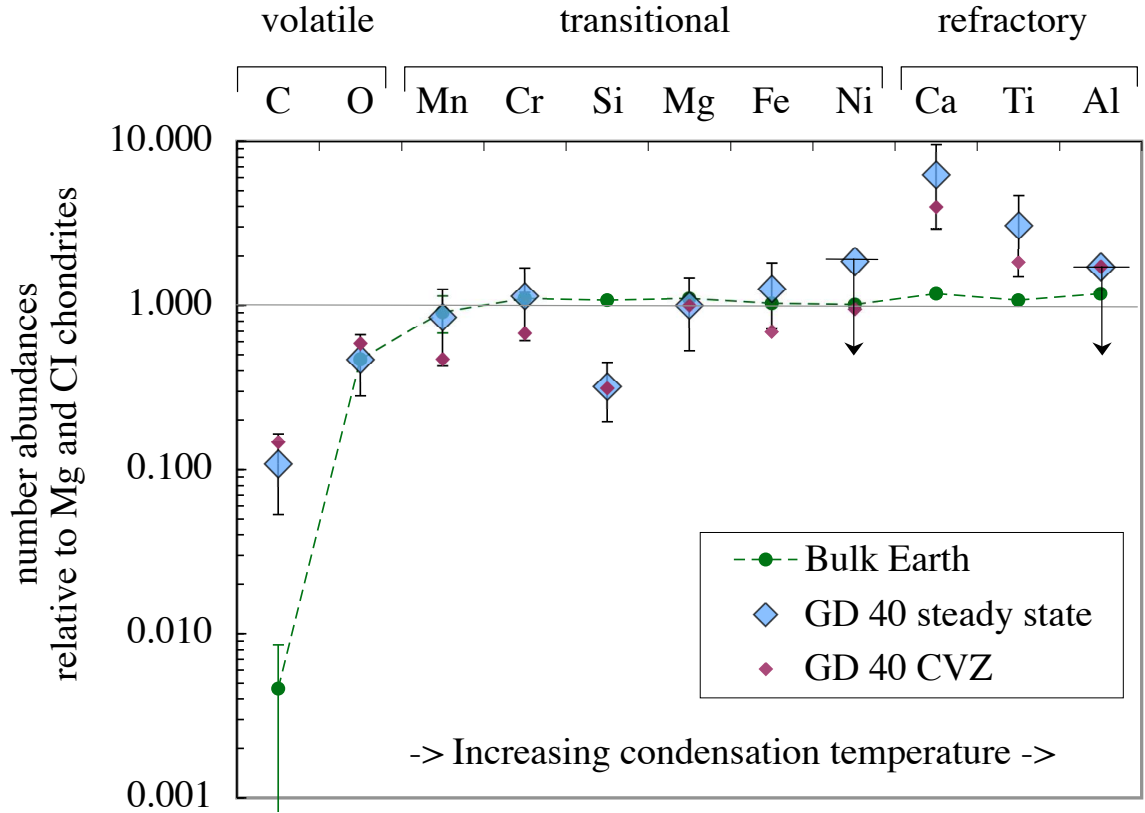}
\caption{Abundances shown normalized to Mg and CI chondrite composition to emphasize the alteration relative to primitive abundances.  Ni and Al are upper limits.    O, Mg, Fe, Cr, and Mn are consistent with bulk Earth abundances.  Similar to Earth, the volatiles C, O are depleted from primitive values.  The refractory elements Ca and Ti are enhanced in the body that pollutes GD 40.  The trend of increasing abundances with condensation temperature most likely is a result of thermal processing.  The depletion of silicon is distinctive and suggests that some kind of differential processing is likely to have occurred in the parent body.  The GD 40 data are from Table 5 (CVZ = convective zone $\equiv$ early-phase); Earth from Allegre et al. (1995) with Earth carbon data from Lodders \& Fegley (1998); CI chondrites from Lodders (2003).}
\label{default}
\end{center}
\end{figure}

\begin{figure}[htbp]
\begin{center}
\plotone{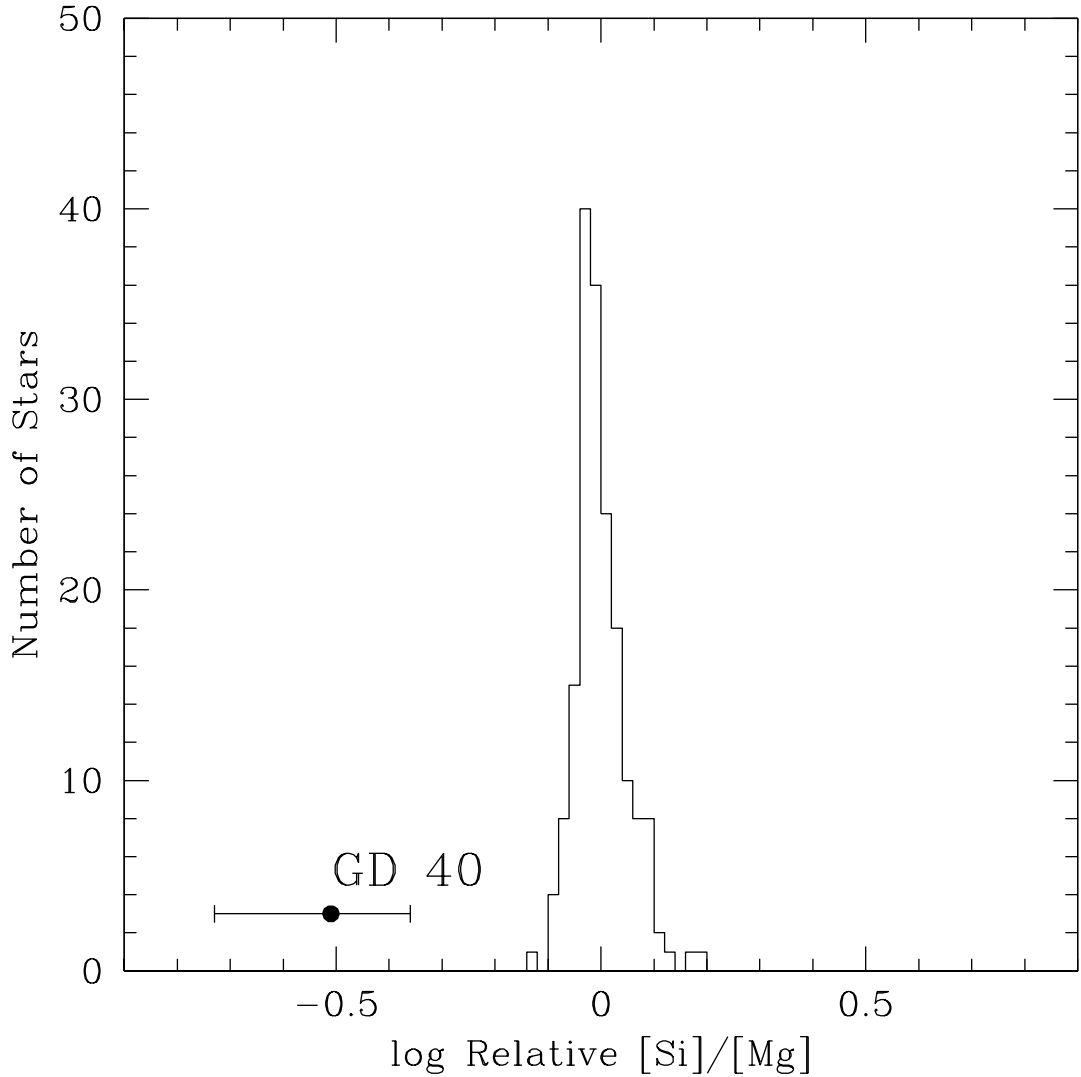}
\caption{Data for 180 F and G dwarf stars from Reddy et al.$\;$(2003).  For a sample of 6 early B-type stars, Przybilla et al. (2008) obtain log [Si]/[Mg] = -0.06 $\pm$ 0.05, within the F and G dwarf distribution.  The GD 40 star+planetary system was probably not born with the distinctive Si/Mg ratio observed in the accreted material.   One possibility is that the GD 40 parent body experienced a substantial differentiation, such that silicon was concentrated in a crust, and Mg in a mantle, and that the crust has been subsequently lost. }
\label{default}
\end{center}
\end{figure}

\end{document}